# Evidence for Wide-Spread AGN Driven Outflows in the Most Massive z~1-2 Star Forming Galaxies[1]

R.Genzel[1,2,3], N.M. Förster Schreiber[1], D.Rosario[1], P.Lang[1], D.Lutz[1], E.Wisnioski[1], E.Wuyts[1], S.Wuyts[1], K.Bandara[1], R.Bender[1,4], S.Berta[1], J.Kurk[1], J.T. Mendel[1], L.J.Tacconi[1], D.Wilman[1,4], A. Beifiori[1,4], G. Brammer[5], A.Burkert[1,4], P.Buschkamp[1], J. Chan[1], C.M. Carollo[6], R.Davies[1], F.Eisenhauer[1], M. Fabricius[1], M. Fossati[4,1], M.Kriek[3], S. Kulkarni[1], S.J. Lilly[6], C. Mancini[7], I.Momcheva[8], T.Naab[9], E.J. Nelson[8], A. Renzini[7], R.Saglia[1,4], R.M. Sharples[10], A.Sternberg[11], S.Tacchella[6] & P. van Dokkum[8]

[1]*Max-Planck-Institut für extraterrestrische Physik (MPE), Giessenbachstr.1, 85748 Garching, Germany (forster@mpe.mpg.de, genzel@mpe.mpg.de)*

[2]*Department of Physics, Le Conte Hall, University of California, 94720 Berkeley, USA*

[3]*Department of Astronomy, Hearst Field Annex, University of California, Berkeley, 94720, USA*

[4]*Universitäts-Sternwarte Ludwig-Maximilians-Universität (USM), Scheinerstr. 1, München, D-81679, Germany*

[5] *Space Telescope Science Institute, Baltimore, MD 21218, USA*

[6] *Institute of Astronomy, Department of Physics, Eidgenössische Technische Hochschule, ETH Zürich, CH-8093, Switzerland*

[7] *Osservatorio Astronomico di Padova, Vicolo dell'Osservatorio 5, Padova, I-35122, Italy*

[1] Based on observations obtained at the Very Large Telescope (VLT) of the European Southern Observatory (ESO), Paranal, Chile (ESO program IDs 073.B-9018, 074.A-9011, 075.A-0466, 076.A-0527, 078.A-0660, 079.A-0341, 080.A-0330, 080.A-0339, 080.A-0635, 081.A-0672, 082.A-0396, 183.A-0781, 087.A-0081, 088.A-0202, 088.A-0209, 091.A-0126, 092.A-0082, 092.A-0091, 093.A-0079). Also based on observations at the Large Binocular Telescope (LBT) on Mt. Graham in Arizona.

[8] *Department of Astronomy, Yale University, P.O. Box 208101, New Haven, CT 06520-810, USA*

[9] *Max-Planck Institute for Astrophysics, Karl Schwarzschildstrasse 1, D-85748 Garching, Germany*

[10] *Department of Physics, Durham University, Science Laboratories, South Road Durham DH1 3LE, UK*

[11] *School of Physics and Astronomy, Tel Aviv University, Tel Aviv 69978, Israel*

# ABSTRACT

In this paper we follow up on our previous detection of nuclear ionized outflows in the most massive ($\log(M_*/M_\odot) \geq 10.9$) z ~ 1-3 star-forming galaxies (Förster Schreiber et al.), by increasing the sample size by a factor of six (to 44 galaxies above $\log(M_*/M_\odot) \geq 10.9$) from a combination of the SINS/zC-SINF, LUCI, GNIRS, and KMOS$^{3D}$ spectroscopic surveys. We find a fairly sharp onset of the incidence of broad nuclear emission (FWHM in the Hα, [NII], and [SII] lines ~450 - 5300 km/s), with large [NII]/Hα ratios, above $\log(M_*/M_\odot)$~10.9, with about two thirds of the galaxies in this mass range exhibiting this component. Broad nuclear components near and above the Schechter mass are similarly prevalent above and below the main sequence of star-forming galaxies, and at z~1 and ~2. The line ratios of the nuclear component are fit by excitation from active galactic nuclei (AGN), or by a combination of shocks and photoionization. The incidence of the most massive galaxies with broad nuclear components is at least as large as that of AGNs identified by X-ray, optical, infrared or radio indicators. The mass loading of the nuclear outflows is near unity. Our findings provide compelling evidence for powerful, high-duty cycle, AGN-driven outflows near the Schechter mass, and acting across the peak of cosmic galaxy formation.

# 1. Introduction

Throughout the last 10 billion years galaxies have been fairly inefficient in incorporating the cosmic baryons available to them into their stellar components. At a halo mass near $10^{12}$ $M_\odot$ this baryon fraction is only about 20% (of the cosmic baryon abundance), and the efficiency drops to even lower values on either side of this mass (e.g., Madau et al. 1996; Baldry et al. 2008, Conroy & Wechsler 2009, Guo et al. 2010, Moster et al. 2010, 2013, Behroozi et al. 2013). Galactic winds driven by supernovae and massive stars have long been proposed to explain the low baryon content of halos much below $\log(M_h/M_\odot)\sim12$ (e.g. Dekel & Silk 1986, Efstathiou 2000). The decreasing efficiency of galaxy formation above $\log(M_h/M_\odot)\sim12$ may be caused by less efficient cooling and accretion of baryons in massive halos (Rees & Ostriker 1977, Dekel & Birnboim 2006). Alternatively or additionally efficient outflows driven by accreting massive black holes may quench star formation at the high mass tail, at and above the Schechter stellar mass, $M_S\sim10^{10.9}$ $M_\odot$ (di Matteo, Springel & Hernquist 2005, Croton et al. 2006, Bower et al. 2006, Hopkins et al. 2006, Cattaneo et al. 2007, Somerville et al. 2008, Fabian 2012).

In the local Universe, such 'AGN feedback' has been observed in the so called 'radio mode' in central cluster galaxies driving jets into the intra-cluster medium (Heckman & Best 2014, McNamara & Nulsen 2007, Fabian 2012), in ionized winds from Seyfert 2 AGNs (e.g. Cecil, Bland, & Tully 1990, Veilleux, Cecil & Bland-Hawthorn 2005, Westmoquette et al. 2012, Rupke & Veilleux 2013, Harrison et al. 2014), and in powerful neutral and ionized gas outflows from buried AGNs in late stage, gas rich mergers (Fischer et al. 2010, Feruglio et al. 2010, Sturm et al. 2011, Rupke & Veilleux 2013, Veilleux et al. 2013, Arribas et al. 2014).

At high-z AGN feedback has been observed in the so called ‘quasar mode’ in broad absorption line quasars (Arav et al. 2001, 2008, 2013, Korista et al. 2008), in type 2 AGN (Alexander et al. 2010, Nesvadba et al. 2011, Cano Díaz et al. 2012, Harrison et al. 2012), and in radio galaxies (Nesvadba et al.2008). However, luminous AGNs near the Eddington limit are rare. Luminous QSOs constitute <1% of the star forming population in the same mass range (e.g. Boyle et al. 2000). QSOs have short lifetimes relative to the Hubble time ($t_{QSO}$~ $10^7 – 10^8$ yr << $t_H$, Martini 2004) and thus low duty cycles compared to galactic star formation processes ($t_{SF}$~ $10^9$ yr, Hickox et al. 2014). It is thus not clear whether the radiatively efficient ‘quasar mode’ can have much effect in regulating galaxy growth and star formation shutdown, as postulated in the theoretical work cited above (Heckman 2010, Fabian 2012).

From deep SINFONI adaptive optics assisted (AO) observations at the ESO VLT, Förster Schreiber et al. (2014a, henceforth FS14a) have recently reported the discovery of broad ionized gas emission associated with the nuclear regions of very massive ($\log(M_*/M_\odot)$>10.9) z~2 ***main-sequence star forming galaxies*** (SFGs) observed as part of the SINS/zC-SINF surveys (Förster Schreiber et al. 2009, and 2014b in preparation, Mancini et al. 2011). For the seven galaxies with best data quality enabling a quantitative analysis, all exhibit

- a very broad, centrally concentrated emission component with FWHM >1000 km/s in the Hα and [NII] (and probably the [SII] λλ 6716/6731) lines, which coincides with the location of a massive stellar bulge revealed by Hubble Space Telescope (HST) near-IR imaging. In several galaxies this broad component is resolved by the AO observations, indicating an intrinsic FWHM diameter of 2 – 3 kpc,

- a (circum)-nuclear ratio of the narrow emission component [NII]/Hα line fluxes of 0.5-0.8, at or above the limit of normal stellar photoionized HII regions, and akin to type 2 AGNs.

The fact that the broad emission component is present in the forbidden [NII] lines as well as its kpc-size extent excludes that the broad emission comes from a virialized, parsec-scale AGN ***broad-line region (BLR)*** in these cases. If so, the >1000 km/s velocity range on kiloparsec scales implies that the broad component cannot be gravitationally bound and must represent a circum-nuclear outflow in the kpc- scale 'narrow-line region' (Netzer 2013). The substantial flux ratio of $F(H\alpha_{broad})/F(H\alpha_{narrow})$~0.3-1 found in these galaxies then suggests the mass loading of these nuclear outflows is substantial ($dM_{out}/dt/SFR$~1, FS14a).

Based on X-ray and mid-infrared indicators, AGN incidence at z>1 increases from a few percent at $\log(M_*/M_\odot)$ ~ 10 – 10.5 up to ~ 15% - 30% at $\log(M_*/M_\odot) > 11$ (e.g., Reddy et al. 2005, Papovich et al. 2006, Daddi et al. 2007, Brusa et al. 2009, 2014, Hainline et al. 2012, Bongiorno et al. 2012). Herschel studies have revealed that the AGN host population is mainly drawn from normal main-sequence SFGs (e.g., Mullaney et al. 2012; Rosario et al. 2012; 2013a). As such, the identification of AGN driven outflows in high mass SFGs may not come as a surprise in a qualitative sense. The tantalizing new and exciting element in FS14a is the possible identification of a nuclear ionized outflow component ***in a large fraction*** of such massive, star forming hosts that may be driven by a central AGN. However, the small size of the FS14a sample prevents any firm conclusion on the incidence and properties of the detected nuclear outflows, although an inspection of various other z~2 small galaxy samples in the literature (Erb et al. 2006, Kriek et al. 2007, Swinbank et al. 2012, as discussed in FS14a) are consistent with a fairly large incidence.

In this paper, we have followed up on these results and present a much larger sample compared to the SINS/zC-SINF sample of FS14a, which includes in particular six times more galaxies at $\log(M_*/M_\odot) \geq 10.9$. We combine the samples from the SINS and zC-SINF surveys (Förster Schreiber et al. 2009, 2014b, Mancini et al. 2011) with SINFONI (Eisenhauer et al. 2003, Bonnet et al. 2004), together with first epoch data from our KMOS$^{3D}$ survey of mass-selected SFGs at $0.7 < z < 2.7$ (Wisnioski et al. 2014) obtained with the new KMOS near-IR multi-IFU instrument on the VLT (Sharples et al. 2012, 2013), massive z ~ 1.5 – 2.5 SFGs from our ongoing spectroscopic survey with the LUCI near-IR multi-object spectrograph at the Large Binocular Telescope (LBT, E.Wuyts et al. 2014a ), and massive z ~ 2 – 2.5 SFGs from the K-band selected near-IR spectroscopic sample of Kriek et al. (2007) observed with SINFONI and with GNIRS at Gemini South. With significantly improved statistics, a wider coverage in specific star formation rate (sSFR) and in redshift, the sample studied here allows us to substantially strengthen our previous findings about the onset and properties of nuclear AGN-driven outflows above the Schechter mass, and to explore trends with redshift and with location of galaxies above or below the main sequence of SFGs.

Throughout, we adopt a Chabrier (2003) stellar initial mass function and a ΛCDM cosmology with $H_0 = 70$ km s$^{-1}$ and $\Omega_m = 0.3$.

# 2. Observations

## 2.1 Data Sets

For the analysis in this paper we included a total of 110 SFGs at z ~ 1-3 with near-IR integral field or slit spectroscopy covering the Hα+[NII] line emission from surveys carried out with SINFONI, KMOS, LUCI, and GNIRS. The targets for these surveys were originally drawn from rest-frame optical, UV, and near-IR selected samples in broad-band imaging surveys with optical spectroscopic redshifts, and from stellar mass-selected samples with near-IR or optical spectroscopic redshifts. Global stellar properties for all the galaxies were derived following similar procedures as outlined by Wuyts et al. (2011b). In brief, stellar masses were obtained from fitting the rest-UV to near-IR spectral energy distributions (SEDs) with Bruzual & Charlot (2003) population synthesis models, the Calzetti et al. (2000) reddening law, a solar metallicity, and a range of star formation histories (including constant SFR and exponentially declining SFRs with varying e-folding timescales). SFRs were obtained from the same SED fits or, for objects observed and detected in at least one of the mid- to far-IR (24μm to 160μm) bands with the Spitzer/MIPS and Herschel/PACS instruments, from rest-UV+IR luminosities through the Herschel-calibrated ladder of SFR indicators of Wuyts et al. (2011b). Details of the derivations are given in the references below; we note that the methods and model assumptions were similar for the different sub-samples (and we corrected the $M_*$ and SFR estimates to our adopted Chabrier IMF when necessary), ensuring consistency for the present study.

Of the full near-IR spectroscopic samples considered, we retained the 110 objects that have the high quality and signal-to-noise ratio (line detections with SNR>10) spectra required for our analysis and that do not have strong contamination by atmospheric OH sky emission around the Hα+[NII] complex. The galaxies have redshifts between z = 0.8 and 2.6 and stellar masses in the range $\log(M_*/M_\odot)$ = 9.4 to 11.7. Most are spatially-resolved in their Hα+[NII] line emission. The sample consists of the following subsets,

1) 33 SFGs with $\log(M_*/M_\odot)$= 9.4-11.5 from the z~1.5-2.5 SINS/zC-SINF survey (Förster Schreiber et al. 2009, Mancini et al. 2011); all but four of these galaxies were observed in AO mode resulting in a typical 0.2”-0.3” FWHM resolution. The four SFGs observed only in seeing limited mode (0.5”-0.6” FWHM resolution) were either well resolved at that resolution (3 cases), or strongly dominated by the nuclear region (1 case). In two large and well resolved SFGs we combined AO and seeing limited data sets to further improve the SNR of the spectra;

2) 56 galaxies with $\log(M_*/M_\odot)$= 10.0 – 11.7 at z = 0.8 – 1.1 and z = 2 – 2.6 observed in natural seeing with KMOS during commissioning and the first year of our KMOS$^{3D}$ survey (Wisnioski et al. 2014, in preparation) carried out as part of guaranteed time observations (GTO). These galaxies form a subset of the total of 210 targets observed (and 174 detected in Hα) so far[2] emphasizing the massive part of the sample: they include (i) all targets at $\log(M_*/M_\odot)$ > 10.6 with emission line detections and (ii) the subset of targets at $\log(M_*/M_\odot)$ < 10.6 that are sufficiently well resolved and exhibit evidence of rotation in their kinematic maps;

---

[2] KMOS$^{3D}$ is a multi-year survey; the current sample includes a fraction of targets for which only part of the planned integration time has been obtained and which will be further observed in subsequent semesters.

3) 10 SFGs at z = 1.5 – 2.5 with $\log(M_*/M_\odot) > 10.6$ from our LUCI multi-object slit spectroscopic survey in natural seeing at the LBT (E.Wuyts et al. 2014a). This LUCI sample includes the large $\log(M_*/M_\odot)$=11.0 SFG EGS-13011166 observed in CO molecular line emission as part of the "PHIBSS1" survey of Tacconi et al. (2013), and for which we obtained high quality spatially-resolved Hα+[NII] emission from slit mapping with LUCI (~0.6" FWHM resolution; Genzel et al. 2013);

4) 1 $\log(M_*/M_\odot) = 11.5$ lensed main-sequence SFG (J0901+1814, Diehl et al. 2009, Saintonge et al. 2013), for which we obtained deep, seeing limited and AO SINFONI data. The no-AO and AO data were combined together to increase the SNR and, accounting for the lensing magnification, the effective source plane resolution is ~0.1" (E.Wuyts et al. 2014b, in preparation);

5) 10 $\log(M_*/M_\odot) \gtrsim 11$ emission line galaxies from the K-band selected z ~ 2 – 2.5 near-IR spectroscopic sample of Kriek et al. (2007) observed with GNIRS and SINFONI in seeing-limited mode. The SINFONI data alone have too low SNR for our analysis, so we used the combined GNIRS+SINFONI spectra as published by Kriek et al. with the following exception. For one object (SDSS1030-2026), lying a factor of ~ 30 in specific SFR below the z = 2.5 main sequence, we recently obtained SINFONI AO-assisted observations, which clearly confirm the presence of a spatially compact and spectrally broad emission line component.

Ranked by ascending stellar mass into four bins, $\log(M_*/M_\odot)$ = [9.4-10.3], [10.3-10.6], [10.6-10.9], [10.9-11.7], our sample breaks up into 17, 19, 30 and 44 SFGs, respectively. In the two most critical highest mass bins, there are each six times more galaxies as in the set available to FS14a.

The distribution of the final sample in stellar mass versus specific SFR is shown in Figure 1, along with that of the underlying population of mass-selected galaxies from the 3D-HST Treasury survey (Brammer et al. 2012, Skelton et al. 2014) in the same z = 0.8 – 2.6 range. To account for the global evolution of star formation properties of galaxies with cosmic time, the specific SFR of every object is computed and plotted relative to the value of the main sequence at its respective redshift and stellar mass, denoted sSFR/sSFR(ms), adopting the parameterization of Whitaker et al. (2012)[3]. Of the 110 SFGs of our sample, 92 lie within ±0.6 dex of the main sequence; they span two orders of magnitude in stellar mass, and cover approximately homogeneously the mass and specific SFR range of the main sequence above $\log(M_*/M_\odot) \sim 10.3$. Three SFGs are outliers above the main sequence. The remaining 14 galaxies, all from the KMOS$^{3D}$, LUCI and Kriek et al.(2007) samples, extend our coverage to significantly below the main sequence, with 6 of them having very low specific SFRs (<0.06 of the main sequence).

A kinematic classification is possible for the 93 of the 110 SFGs that have IFU data (this includes the Keck/OSIRIS data published by Law et al. 2012 for one of our LUCI targets, Q2343-BX442). In terms of kinematics, 73 of these sources have a ratio of rotation/orbital velocity to intrinsic velocity dispersion $v_{rot}/\sigma_0>1$ and are plausibly rotating disks (see Newman et al. 2013; Wisnioski et al. 2014, in preparation). Although we emphasized objects with evidence for rotation in choosing the lower-mass KMOS$^{3D}$ objects for this study, the high disk fraction is not surprising and consistent with the growing evidence that a majority of massive z ~ 1 – 2.5 SFGs are disks based on kinematic and morphological properties (see also, e.g., Shapiro et al. 2008; Genzel et al. 2008, 2014a; Förster Schreiber et al. 2009; Épinat et al. 2009, 2012; Jones et al. 2010; Wuyts et al. 2011a; Lang et al. 2014). Three sources are identified as candidate minor mergers, and four are candidate major mergers. Seven of the lower mass galaxies ($\log(M_*/M_\odot)<10.4$) show no or little evidence for rotational support and are classified as 'dispersion dominated'. Three compact galaxies in the highest mass bin exhibit little evidence for narrow line emission as expected from star formation activity and are completely dominated by very broad line emission, probably due to a Type I AGN broad line region. These three objects will hereafter be referred to as "candidate BLR sources." Our preferential inclusion of rotating systems among the lower mass KMOS$^{3D}$ targets may tend to emphasize larger galaxies that are more easily resolved in seeing-limited KMOS data, although SINFONI targets with higher resolution AO data dominate at the low- $M_*$ end of the present sample; we return to this point below. These kinematic identifications are listed in column 3 of Table 1, which also summarizes the salient parameters of our sample.

[3] The exact parameterization of the main sequence of SFGs varies among different studies, which is attributed to the impact of different sample selection, survey completeness, methodology applied to derive the stellar masses and SFRs, among other factors. The Whitaker et al. (2012) fits provide a good representation of the locus of SFGs in our comparison 3D-HST sample above $\log(M_*/M_\odot) \sim 10.3$, encompassing our three highest mass bins comprising 85% of our sample. At lower masses, a difference becomes apparent (see Figure 1); an alternative fit to main-sequence SFGs from 3D-HST is beyond the scope of this paper, so we keep the Whitaker et al. parameterization bearing in mind that the quantitative offset from the main sequence of our $\log(M_*/M_\odot) < 10.3$ galaxies could be more uncertain.

We verified that the requirements imposed when selecting the objects for our study do not introduce significant biases that would affect the results of our analysis, in particular the need for an H$\alpha$ detection, the emphasis on high quality and SNR data sets, and the preferential inclusion of better resolved objects towards lower masses from KMOS$^{3D}$. To this aim, we considered the sSFR and size distributions in the $M_*$ bins defined above of all objects from the parent KMOS$^{3D}$, SINS/zC-SINF, LUCI, and GNIRS+SINFONI near-IR spectroscopic samples, and of the underlying population of SFGs in the same stellar mass and redshift ranges (taken from the 3D-HST survey, and defined as having an inverse sSFR greater than three times the Hubble time at their redshift). For the sizes, we used the major axis effective radius measured from HST *H*-band imaging, available for > 90% of the objects in the near-IR spectroscopic samples and the 3D-HST survey (Table 1, and also van Dokkum et al. 2008; Kriek et al. 2009; Förster Schreiber et al. 2011; Lang et al. 2014; van der Wel et al. 2014; Tacchella et al. 2014).

Altogether, the fractions of H$\alpha$-detected objects among the full parent near-IR spectroscopic samples are ~ 80% - 90% in the three lowest $M_*$ bins. The H$\alpha$ detection fraction drops to ~ 65% in the highest $M_*$ bin, which is largely driven by the fact that we also included objects well below the main sequence (i.e., at very low sSFRs) in our observations. There is a trend of somewhat lower detection fractions for objects below the main sequence or with sizes smaller than the median over all SFGs (from ~ 90% to ~ 60% between low- and high- $M_*$ bins, compared to ~ 90% to 75% for objects above the main sequence or with sizes larger than the median for SFGs), again driven by targets with low sSFR/sSFR(ms) < 0.1 that also tend to be more compact (e.g., van der Wel et al. 2014). These detection fractions and trends are essentially the same when considering only the KMOS$^{3D}$ targets observed so far and with their current integration times.

In terms of range and median values, the sSFR/sSFR(ms) and size distributions of the underlying SFG population are overall well covered by the parent near-IR spectroscopic samples as well as by the Hα-detected subsets and the objects included in the present study. The most significant differences are as follows. In the lowest $M_*$ bin, the parent near-IR spectroscopic samples preferentially probe the part of the SFG population with higher sSFR/sSFR(ms) and larger sizes, by factors of around 3 and 1.8 in the median (due in part to their $M_*$ distribution weighted towards the more massive objects compared to the bulk of SFGs in that $M_*$ interval). The same trend applies to the Hα-detected subset and to the objects analyzed in this paper. At $\log(M_*/M_\odot) < 10.3$, our sample is largely dominated by SINS/zC-SINF galaxies at $z \sim 1.5 – 2.5$ with AO-assisted SINFONI observations (Table 1), for which the typically 3 – 4 times higher resolution compared to seeing-limited data helps to better resolve smaller objects (see also Newman et al. 2013). Towards higher masses, the SFG population is well covered and, in addition, the objects from the KMOS$^{3D}$ and GNIRS+SINFONI parent samples extend to lower sSFR/sSFR(ms) and smaller sizes than the bulk of SFGs, by design of these surveys (*K*-band selection with no SFR cut for the GNIRS+SINFONI sample, $M_*$ selection with very low SFR < 1 $M_*$/yr cut and typically long integrations for KMOS$^{3D}$). The median sSFR/sSFR(ms) and sizes are ≈ 1.7 times lower than for SFGs in the same $M_*$ interval. A similar trend is seen among the Hα-detected subset and for the objects included in the present work, although with smaller differences relative to the SFG population.

To summarize, the high Hα detection rate of the full KMOS$^{3D}$, SINS/zC-SINF, LUCI, and GNIRS+SINFONI samples, and the similarity in ranges and median properties (sSFR, size) of the Hα-detected objects as well as of those entering the sample studied here compared to the underlying population of SFGs, indicate that our sample probes well the SFG population at similar redshift and above $\log(M_*/M_\odot) > 10.3$. In the lowest $M_*$ bin, our sample preferentially includes objects towards larger sizes and higher sSFRs but this bias is unlikely to affect the main findings about the changes in emission line profile and outflow properties discussed in the following Sections, which occur around $\log(M_*/M_\odot) \sim 10.9$ and are thus well enough sampled by the three higher $M_*$ bins. When including the population of massive galaxies well below the main sequence of SFGs, into the regime of quenching/quiescent galaxies, the Hα detection fractions drop most significantly (though they are still around ~60% in the highest $M_*$ bin) and our sample may not yet probe the bulk of that population in terms of emission line properties -- unsurprisingly given the nature and very low SFRs of these galaxies.

### 2.2 Data Analysis

The observations and data reduction procedures are presented by Förster Schreiber et al. (2009; 2014b in preparation) for the SINS/zC-SINF SINFONI data, by Wisnioski et al. (2014, in preparation) and Davies et al. (2013) for the KMOS data, by E.Wuyts et al. (2014a; 2014b in preparation) for the LUCI sample and the SINFONI data of the lensed J0901+1814, and by Kriek et al. (2007) for the GNIRS+SINFONI data, to which we refer the reader for details. We focus here on the analysis of the reduced data.

For the SINFONI and KMOS data sets from the SINS/zC-SINF and KMOS$^{3D}$ surveys, the SINFONI observations of J0901+1814 and SDSS1030-2026, and the LUCI slit-mapping data of EGS13011166, we followed the methodology of Shapiro et al.

(2009), Genzel et al. (2011), Newman et al. (2012) and FS14a.The fully reduced data cubes were first median-subtracted (to remove continuum emission, which is well detected in most of the more massive SFGs of our sample), and 4-σ-clipped blue- and red-ward of the Hα+[NII] emission complex to remove OH sky emission line shot noise. In a few cases where an OH sky line was very close to the narrow (star formation-dominated) Hα emission, we interpolated over one to three spectral channels to remove the OH noise. The cubes were then spatially smoothed with a Gaussian of FWHM between 2 to 4 pixels (depending on SNR, source and beam size), and then a single Gaussian line profile was fitted for each pixel to extract a smoothed velocity field of the galaxy. This velocity field was then applied in reverse to the original data cube to remove large scale velocity gradients from orbital motions. This technique minimizes the impact of velocity broadening due to orbital motions in the final extracted spectra, and at the same time improves the SNR for detecting faint features and line wings. The method is somewhat questionable in compact sources with unresolved strong velocity gradients, as it cannot then remove the gradients, which instead result in increased central velocity dispersions.

From the velocity-shifted cube for each galaxy we extracted a spectrum in an aperture of diameter ~0.3”-0.4” (for AO data with 0.05” pixels) to 0.6” (for seeing limited data with 0.125-0.2” pixels) centered on the kinematic centroid, which coincides with the continuum peak for almost all of the SFGs in the highest mass bins. For galaxies in the two lower mass bins, there is often no or only a weak nuclear concentration of continuum light, consistent with the lower bulge to disk ratios found based on high resolution HST imaging of these sources (Lang et al. 2014, Tacchella et al. 2014). The above aperture sizes correspond to physical radii of ~1.1 – 1.6 kpc (AO data) to 2.2 – 2.4 kpc (seeing-limited data) at the redshifts of our galaxies. For simplicity, throughout the paper we will refer to these spectra as "nuclear spectra" although they cover the nuclear and circum-nuclear emission of the galaxies. We also extracted outer "disk spectra" outside the nuclear aperture, over a region with significant Hα emission. The final nuclear and disk spectra for each galaxy were normalized to a peak amplitude at Hα of unity and interpolated onto a common velocity sampling of 30 km/s.

The quality of the spectra extracted from the data cubes above is good to excellent, owing to on-source integration times varying between 2 and 23 h, with an average and median of about 8 hours. The median SNR per spectral element of the nuclear and disk spectra is ~10.

For the slit spectroscopy obtained with LUCI, and the published GNIRS+SINFONI data of Kriek et al. (2007), we used the source-integrated spectra as proxies of the nuclear emission. Whereas this choice implies a potentially larger contribution from the disk regions to the nuclear spectra, inspection of the two-dimensional LUCI slit spectra and of the SINFONI Hα maps of Kriek et al. (2007) indicates that the bulk of the line emission originates from the central regions. The impact on the co-added spectra discussed below and in subsequent sections is, however, small since these 18 LUCI and GNIRS+SINFONI spectra represent only ≈15% of all our data sets (or 20% and 27% in the two highest mass bins), and because of their typically lower than average SNR they are substantially down-weighted in the co-adding (see below).

In constructing the various co-added spectra we used two approaches. In one approach, we gave all galaxies the same statistical weight, but left out a few lower-SNR galaxies in the sample. This choice obviously does not optimize the SNR of the co-added spectrum but instead yields the most likely ‘average’ spectrum of the chosen sub-sample, and is least affected by outliers. In the second approach we gave each galaxy a weight proportional to its signal to noise ratio, to generate the best quality co-added spectrum. We did not pursue a weighting proportional to $SNR^2$, as this would have given overly strong emphasis to a few galaxies with the best SNR. We also compared results by splitting up the sub-sample comparing their properties. We find that these different methodologies make little difference in the resulting spectra, demonstrating that the properties of our co-added spectra, at least for sub-samples of 5 to 10 galaxies, are robust. For these reasons we chose in the end, for the display of co-added spectra and quantitative analyses the SNR-weighting scheme (with one exception, see Section 3.1). The final co-added spectra were re-binned to 40 km/s, roughly representing two samples per average intrinsic instrumental FWHM resolution of SINFONI, KMOS and GNIRS.

Motivated by the earlier analysis of Genzel et al. (2011), we used multiple Gaussian fitting for the spectral analysis, with the following input assumptions,

- the systemic velocities and widths of the narrow Hα, [NII] and [SII] line components are the same, and likewise for the broad components,
- the ratio of [NII] λ6548/λ6583 is 0.32 (Storey & Zeippen 2000),
- the flux ratio [SII] λ6716/λ6731 in the broad component (if detected) is ~1, similar to that found in the narrow component in almost all of our SFGs and near the low-density limit.

This leaves then the following free fitting parameters: the FWHM line widths of the narrow and the broad components ($\Delta v_{narrow}$, $\Delta v_{broad}$), the velocity shift between their centroids ($\delta v_{broad}$), the flux ratios [NII] $\lambda$6583 / H$\alpha$ in the narrow and broad components and H$\alpha_{broad}$/H$\alpha_{narrow}$, and, in cases where the [SII] lines were fitted as well, the flux ratios [SII] $\lambda 6716_{narrow}$/H$\alpha_{narrow}$, [SII] $\lambda 6716_{narrow}$/[SII] $\lambda 6731_{narrow,}$ and [SII] $\lambda 6716_{broad}$/H$\alpha_{narrow}$. All narrow and broad Gaussian components were always fit simultaneously.

As will be seen from the discussion below (see also Genzel et al. 2011), the assumption of Gaussian line shapes is well justified for the narrow component (in terms of the central limit theorem of many individual HII regions contributing to the final shape where large velocity gradients have been removed). This justification is less obvious for the broad component, which in some cases appears to exhibit a blue/red asymmetry, in which case the inferred line widths serve as a first order description. When splitting the sample into more numerous, and smaller sub-samples, or analyzing the lowest mass bin, the SNR of the broad emission can become marginal for quantitative fitting of its width. In this case, and motivated by the fairly constant velocity width of the broad component in the disk and lower mass bins (see also Newman et al. 2012), we adopted $\Delta v_{broad}$=380 km/s as a fixed input parameter. For the faintest low-mass galaxies with weak [NII] emission, we also assumed that the [NII] $\lambda$6583/H$\alpha$ flux ratio in the broad component was twice that in the narrow component, motivated by the findings at higher masses.

# 3. Results

The discovery observations of FS14a raised three key issues we wish to explore in this paper. How common are the (circum)-nuclear ionized outflows? How do their outflow rates and outflow velocities vary with location of the SFG in the stellar mass – specific SFR plane, and with redshift? What drives and excites these outflows, AGNs or (circum)-nuclear starbursts?

Tackling these questions requires a much larger sample of galaxies than was available to FS14a, and is now possible with the new high quality AO and seeing limited data sets assembled in this paper, comprising 110 SFGs with $\log(M_*/M_\odot)$=9.4-11.7. In particular, in the two highest mass bins this sample increases the data set used by FS14a from 13 to 74 SFGs. The extended sample also covers the distribution of the main sequence of SFGs in the $\log M_*$-sSFR plane more homogeneously (especially at $\log(M_*/M_\odot) > 10.3$) and pushes the coverage at the highest masses to specific SFRs significantly below the main sequence as can be seen in Figure 1 (see also Table 1). Moreover, our new sample includes 29 SFGs at z = 0.8 – 1.6 and 81 at z = 2 – 2.5 (blue circles and red squares in Figure 1), allowing us to investigate the frequency and properties of nuclear outflows at lower redshifts compared to the FS14a study.

## 3.1 Detection of broad nuclear components

In 34 of the 110 SFGs of our sample we detect a significant broad component in their ***individual nuclear*** spectra; the spectra of these 34 SFGs are plotted in Figure 2. We identify a ‘broad component’ detection when the broad component emission flux is significant based on the uncertainties from the multiple simultaneous Gaussian fits (described in Section 2.2). We have shown previously that the assumption of Gaussian line profiles (of typically FWHM ~140 km/s) is empirically well justified for individual giant star forming clumps (Genzel et al. 2011). After removal of large scale velocity gradients, our spatially-resolved SINFONI and KMOS data show that also the galaxy wide spectra are near Gaussian with FWHM line widths ranging between 150 and 320 km/s. The underlying excess broad components in the spectra of Figure 2 have FWHM ranging from 430 to 5300 km/s.

Does such an excess 'broad component' necessarily imply a separate broad component, or could it also be the result of beam-smeared unresolved orbital motions, especially for the seeing limited data sets? The wings of the instrumental spectral profile of the SINFONI instrument are negligible compared to the line widths of the broad (and narrow) components discussed here (Genzel et al. 2011, FS14a). The same can be said about the KMOS and GNIRS instruments. LUCI has a more complex spectral response function but the statement above still holds for the SFGs discussed here. To explore the issue of spatial 'cross talk' we consider an illustrative case of a SFG with a size and mass representative of the disks observed in the SINS/zC-SINF and KMOS$^{3D}$ surveys (Förster Schreiber et al. 2009; Mancini et al. 2011; Newman et al. 2013; Wisnioski et al. 2014, in preparation). We set up an inclined (sin (i)=0.76) exponential disk (of effective radius $R_e$~6.5 kpc), plus bulge model with a total stellar mass of $M_*=1.5\times10^{11}\ M_\odot$, and a fairly flat projected rotation curve of $v_{max}$~240 km/s, which we then convolved with a seeing limited PSF of FWHM 0.55", added appropriate Gaussian noise (comparable to our SINFONI and KMOS data), and assumed [NII]/Hα =0.3. We then analyzed the model data cube in the same manner as for the real data, including the de-shifting of the large scale velocity gradients, extraction of inner and outer disk aperture spectra, etc. The extracted disk spectrum (FWHM ~160 km/s) of this model galaxy is shown in the left panel of Figure 3. The equal weight average of the outer disk spectra of 43 high quality SFGs throughout the full mass range of our sample is shown in blue, and in green is the best fit broad component for that spectrum. It is obvious that beam-smeared orbital motions even in a massive SFG galaxy cannot account for the broad emission in the average outer disk spectra of our sample. For main-sequence SFGs as observed in our SINS/zC-SINF and $KMOS^{3D}$ surveys (or other SFG samples observed with near-IR integral field spectrographs, e.g., Law et al. 2009, Épinat et al. 2009, 2012), this statement is conservative since the orbital motions in most of the galaxies would be smaller than in the massive model system we used. The central 0.3-0.4" diameter aperture spectrum of our model galaxy has a FWHM of 440 km/s. While the beam smearing of unresolved nuclear motions could contribute to, or perhaps even dominate a nuclear width of ~400 km/s in a massive SFG galaxy, it obviously cannot account for ~1000 km/s components we observe for the $\log(M_*/M_\odot)>10.9$ galaxies as described below [4]. The same conclusions apply for the SINFONI+AO data; while the core of the AO PSF has a narrow FWHM $\approx$ 0.2" , it exhibits significant broad wings with a FWHM ~ 0.55" corresponding to the uncorrected seeing (FS14a, Förster Schreiber et al. 2014b, in preparation).

[4] Dense compact quiescent galaxies at z ~ 1 – 3, with stellar masses of $\log(M_*/M_\odot)$ ~ 11 and effective radii ~ 1 kpc have typical stellar velocity dispersions of 300 – 400 km/s (e.g., van Dokkum et al. 2009; Bezanson et al. 2013; van de Sande et al. 2013; Belli et al. 2014). Although we cannot exclude that star-forming progenitors of such very dense "cores" may be present among our galaxies and cause FWHMs up to ~ 100 km/s, these are very rare and unlikely to dominate our sample (e.g., Tacconi et al. 2008; Nelson et al. 2014).

In practice, a broad component can be detected in individual spectra if its integrated flux is at least 10 % of the narrow component, and its width is at least twice that of the narrow component. The average signal to noise ratios of our spectra are comparable across the stellar mass range covered, thus making the same relative broad line fraction as easy or difficult to detect at $\log(M_*/M_\odot)$~10, as at $\log(M_*/M_\odot)$~11.3. We have verified this assessment quantitatively by adding model broad components of FWHM 500 and 1500 km/s in Hα and the [NII] lines in various strengths to the stacked central and outer disk spectra in the different mass bins (leaving out those stacks with strong detected broad components), and then analyzing the spectra in the same manner as described in section 2.2. In these stacks (of typically 8-11 galaxies each) the minimum detectable broad component, in the sense of a significant/correct extraction of its width and flux, is about 15-20% of the narrow component in terms of flux ratio, more or less flat across the mass range sampled by our data and similar for both widths. These detection limits are shown as thick black and magenta lines in the right panel of Figure 3. Weaker broad components (to about 10% of the narrow flux) can still be detected but their inferred properties are uncertain.

In terms of these definitions, a significant intrinsic broad nuclear component is present in each of the 34 SFGs in Figure 2. This broad component obviously varies greatly from source to source in width and strength relative to the narrow Hα and two [NII] lines. We will return to the detailed properties of this broad emission when we analyze the high quality co-added spectra.

In addition to these ‘firm’ detections (labelled as quality ‘1’ or ‘2’ in column 7 of Table 1), there are thirteen ‘candidates’ (labelled as quality ‘0.5’ in column 7 of Table 1) with possible but individually marginal broad nuclear components. Broad components are also detected in the outer 'disks' of a number of our SFGs, as previously discussed in Genzel et al. (2011) and Newman et al. (2012). However, in these cases the extended broad component in Hα and [NII] typically has a FWHM of ~380 km/s, about twice that of the narrow component.

We note that because a majority of the data sets considered here (80 of 110) were obtained in natural seeing (and 18 of them consist of source-integrated spectra), the fraction of galaxies in which we identify a broad nuclear emission component may represent a lower limit. Indeed, due to the more significant effects of beam-smearing in seeing-limited data, broad nuclear emission may be more easily outshined, or diluted, by emission from the disk regions (see also discussion by FS14a).

## 3.2 Spectral properties of the broad nuclear emission

### *3.2.1 Line widths, velocities and flux ratios*

To determine the ***average*** properties of the nuclear emission we ***co-added spectra*** of different sub-samples, keeping in mind the substantial variation of profiles seen in the individual sources in Figure 2. Following FS14a we started by averaging the individual spectra of all 31 galaxies in the highest mass bin at $\log(M_*/M_\odot) \geq 10.9$ that have firm or candidate individual detections of a broad nuclear component (quality criteria 0.5,1 or 2 in column 7 of Table 1). We excluded here (as elsewhere below) the spectra of the candidate BLR sources. We weighted each spectrum by its signal-to-noise-ratio (SNR) given in column 6 of Table 1. This stacked spectrum is shown in the left panel of Figure 4 (grey line), and exhibits a prominent broad emission component (blue line, after subtraction of the narrow emission (grey) from the multi-Gaussian fits to the stacked spectrum as described in Section 2.2) with wings extending to 2000 km/s to the blue and the red relative to the narrow Hα emission.  The corresponding stacked outer disk spectrum of SFGs in the same mass bin, plotted in the middle panel of Figure 4 (grey line), also shows a broad component (blue line), but of much smaller width (FWHM 400-500 km/s in Hα), demonstrating  that the very broad component indeed only occurs on average in the central regions. The multi-Gaussian component fit to the nuclear spectrum shows that the broad nuclear component has a FWHM of 1710±70 km/s in Hα and [NII] λλ 6548/6583 (indicated by the red and green lines in the left panel of Figure 4, respectively). The narrow [NII] λ6583/Hα flux ratio is 0.55±0.02 and the broad to narrow Hα flux ratio is 0.37±0.08. The broad [NII] λ6583/Hα ratio is about five times larger than the narrow ratio, 2.7±0.7; the strong broad [NII] emission thus dominates the overall broad emission component and explains its overall asymmetric shape with a strong redshifted peak and a long blueshifted wing (c.f. FS14a).

The broad emission is also confidently detected at 9σ in the [SII] λλ 6716/6731 lines, as shown in the right panel of Figure 4 (blue line). The broad to narrow [SII] line flux ratio is 0.102 (±0.015), and the ratio of the narrow [SII] λ6716 to Hα ratio is 0.12 (±0.01). However, the exact value of these ratios depends also on the broad flux ratio of [SII] λ6716/λ6731, which cannot be uniquely constrained from the data, and which we assumed to be ~1, motivated by the ratio in the narrow [SII] lines. All these values are summarized in Table 2, are in excellent agreement with FS14a, and are quite robust to the sample selection. Changing the sample to include only the best individual detections of nuclear broad emission, or stacking all 35 SFGs with $\log(M_*/M_\odot)>10.9$, or extending the lower mass limit to 10.6, all yield a broad profile with FWHM ~1300-1800 km/s in each Hα, [NII], and [SII], which is dominated by strong broad [NII] emission.

A possible alternative, and formally also acceptable decomposition of the specific co-added nuclear spectrum in the left panel of Figure 4 is obtained if one assumes that the broad emission is due to Hα only, as would be expected for BLR emission (c.f. Netzer 2013). For the three candidate BLR sources in our sample (GOODSN-07923, COS4-14596 and COS4-21492) this explanation may indeed be fully appropriate. These three sources have the largest broad line widths (FWHM 5300, 5200 and 2500 km/s) and at the same time do not show evidence for narrow (or broad) [NII] or [SII] emission, suggesting that in these cases the line emission is indeed dominated by very dense gas from a classical, virialized BLR very close to the central massive black hole (c.f. Netzer 2013). For the co-added nuclear spectrum in Figure 4, however, the broad emission of FWHM ~2200 km/s would then be redshifted by ~310 km/s relative to the narrow Hα, [NII] and [SII] emission. In this explanation the broad Hα emission would have to come from a BLR in most SFGs entering into the co-added profile. Such a large shift between the broad and narrow Hα lines for most or all galaxies is highly unlikely, when compared to local SDSS AGN results (Bonning, Shields & Salviander 2007, Liu et al. 2014, Mullaney et al. 2013). Probably the most conclusive argument against a BLR explanation for the majority of our sources is the clear detection of a broad [SII] line in the co-added spectrum and in individual sources, with the same width as for the Hα and [NII] lines (right panel of Figure 4), and with a centroid velocity consistent with that of the narrow emission. Of course, for those of our SFGs with spatially resolved broad nuclear emission a BLR explanation is excluded in any case.

Another decomposition with the broad [NII] emission having the same [NII]/Hα ratio as in the narrow component is also possible but is less likely for the nuclear spectrum in Figure 4 (and other stacks discussed below), since the broad [NII] λ 6548 emission is weaker and cannot help explaining the strong blue excess in the wings of the overall broad emission in Figure 4. This then would result in a very asymmetric line profile of the broad emission, as well as a poorer fit to the data (c.f. FS14a).

In summary of this section, we fully confirm in a much larger sample the discovery of FS14a that the most massive near-main sequence SFGs at z~1-3 frequently exhibit a very broad nuclear component that is present in Hα, [NII], and [SII] emission lines, and is much wider than in the outer disk regions of the same galaxies. Combined with the evidence that the broad emission is spatially resolved (FWHM~ 2-3 kpc) in 4-5 of these SFGs (FS14a, E.Wuyts et al. 2014, in prep) and that the broad emission is present in the forbidden lines of [NII] and [SII], we have a compelling case that the broad emission represents a powerful nuclear outflow. The blueshift of the broad Hα emission relative to the narrow emission in Figure 4 (-130 (±40) km/s, second row and second column of Table 2) is also consistent with an outflow interpretation, because of the plausible presence of internal differential extinction (Genzel et al. 2011).

### *3.2.2 Line ratios and constraints on the excitation mechanisms*

We next explore the mechanism(s) exciting the broad nuclear line emission, based on rest-optical diagnostic line ratios (e.g., Baldwin, Phillips, & Terlevich 1981; Veilleux & Osterbrock 1987). Figure 5 shows the line ratio properties derived from the data of our sample and compares them with various recent excitation/ionization models (Kewley et

al. 2001, 2006, 2013; Allen et al. 2008; Rich et al. 2010, 2011; Sharp & Bland-Hawthorn 2010; Newman et al. 2014). There is growing evidence that at z ~ 1 – 2, the physical conditions of the interstellar medium (ISM) of SFGs are different than those of normal SFGs at z ~ 0 (e.g., Steidel et al. 2014). High-z SFGs exhibit an offset towards higher excitation in the classical diagrams plotting [OIII] $\lambda$5007/H$\beta$ versus [NII] $\lambda$6583/H$\alpha$, [SII] $\lambda\lambda$6716+6731/H$\alpha$, and [OI] $\lambda$6300/H$\alpha$, such that the criteria to distinguish pure stellar photoionization from AGN and/or shock excitation devised based on normal z ~ 0 SFGs may not be directly applicable at higher redshift (e.g., Kewley et al. 2013, and references therein). Measurements have been published for [OIII] $\lambda$5007/H$\beta$ versus [NII] $\lambda$6583/H$\alpha$, showing that normal, non-AGN SFGs occupy the region between the locus of normal local SFGs and HII regions, and the theoretical "maximum starburst line" from Kewley et al. (2001), overlapping with the location of nearby starburst systems (e.g., Shapley et al. 2005; Kriek et al. 2007; Liu et al. 2008; Trump et al. 2013; Steidel et al. 2014). As illustrated in the middle left panel of Figure 5, this "extreme starburst line" (thick black curve) coincides well with the upper envelope of pure stellar photoionization models for ISM conditions arguably more appropriate at z ~ 1 – 2. Therefore, we interpret our emission line ratios using the Kewley et al. (2001) extreme starburst line in all three diagnostic diagrams considered here.

As already found by FS14a and confirmed in the spectra of Figure 2 the nuclear spectra in the $\log(M_*/M_\odot)>10.9$ SFGs typically have high (total) [NII] $\lambda$6583/H$\alpha$ ratios (log([NII]/H$\alpha$) ranging from -0.7 to 0.2, see Table 1 and histogram at the top left of Figure 5). The broad component [NII] $\lambda$6583/H$\alpha$ ratios in the stack of Figure 4 and in the best individual broad line sources are even greater (log([NII]/H$\alpha$) ~ 0 – 0.4 ). These ratios are at or above the highest values explainable by stellar photoionization for super-solar metallicity (Veilleux & Osterbrock 1987, Kewley et al. 2001, 2006, 2013). For the same spectra the ratio of narrow [SII] to Hα flux is log([SII] λλ6716+6731/Hα)= -0.57±0.05 (Table 2).

For a small subset of 6 broad emission sources we also detect [OI] λ 6300 (the top right panel of Figure 5 shows the co-added spectrum) with log([OI]/Hα)~ -1 (Table 2). For five sources (GS3-19791, D3a-15504, Q2343-BX610, D3a-6004, GOODSN-07923) we have [OIII] λ5007/Hβ from seeing-limited SINFONI and LUCI observations, with source-integrated values of log([OIII]/Hβ ) between +0.25 and +0.75 (Newman et al. 2014). Because of the beam smearing, the source-integrated ratios are probably lower bounds to the nuclear [OIII]/Hβ ratios (see also FS14a).

In the diagnostic diagrams of Figure 5, the galaxies with several line ratios, as well as their averages, overall occupy the area at and above the extreme 'starburst' line of Kewley et al. (2001), where the narrow line regions of metal rich AGN are observed to be located in the local universe, and expected to lie at higher z (Kewley et al. 2013). Of those only Q2343-BX610 could be due to pure stellar photoionization. The narrow emission of GOODSN-07923 is fully consistent with stellar photoionization but its broad emission almost certainly is due to a BLR. Combining the constraints, the alternative of pure shock excitation (Dopita & Sutherland 1995; Allen et al. 2008; Rich et al. 2010; 2011; Sharp & Bland-Hawthorn 2010) also seems unlikely in these cases, with the exception of D3a-6004. For the other nuclear broad emission SFGs for which we only have [NII]/Hα (top histogram in Figure 5), the high values also are in agreement with the best cases discussed above and favour the AGN excitation (and/or shock excitation) explanation.

Figure 5 provides convincing evidence that for those of our SFGs for which multiple line ratios are available the observed line ratios are consistent with a significant AGN contribution to the gas excitation. However, when allowing also the combination of different mechanisms it is possible to explain the observed line ratios with metal rich gas, ionized and excited by a combination of fast shocks and stellar radiation, in agreement with Newman et al. (2014). This possibility is indicated by the grey thick arrows in the diagnostic diagrams of Figure 5. As discussed by Newman et al. (2014) and FS14a, mixed contributions of different excitation mechanisms to the observed line emission could partly be attributed to beam-smearing, since even for our best resolution SINFONI+AO data, the smallest spatial scales probed are around 1 – 2 kpc.

### 3.3 Incidence and properties of nuclear broad components as a function of mass, specific star formation rate and redshift

In this section, we explore trends in the broad component emission as a function of galaxy properties and redshift. To this aim, we consider the fraction of objects with detected broad nuclear emission as a function of stellar mass, offset from the main sequence in SFR, and bulge mass, shown in Figure 6. We also derive the line profile properties of the broad component from co-added spectra of galaxies in different bins of stellar mass, sSFR, and redshift. These spectra are plotted in Figures 7 and 8, and the derived trends are shown in Figure 9. Again, the three candidate BLR sources are excluded from the analysis.

### *3.3.1 Correlation of the broad nuclear components with galaxy stellar mass*

Inspection of Figure 1 shows that the ***individual*** firm and candidate detections of broad nuclear components cluster in the two high mass bins. This is demonstrated more quantitatively in the histogram distributions in Figure 6 (middle panel) and summarized in Table 3. Below $\log(M_*/M_\odot)=10.3$ none of the individual SFGs shows such a broad nuclear component, and there are not even any possible candidates. Between $10.3< \log(M_*/M_\odot)<10.9$ the incidence of a broad nuclear component in the individual spectra is between 20 and 26 (±10)%, depending on whether or not SFGs with candidate detections are included. Then above $\log(M_*/M_\odot)=10.9$, 55 (±11)% of objects show a firmly detected broad nuclear component of FWHM ~500-5200 km/s, where the quoted error bars (here and below) are the Poisson uncertainties. If the broad emission candidate sources are included, the incidence increases to 77 (±13) %. While the quality of their individual spectra is not sufficient to classify the latter reliably, a weighted co-add of the spectra of the 10 candidates in this mass range exhibits the same properties as those of the SFGs with firmly detected broad component emission: a broad FWHM of 610 km/s in Hα and [NII], and a narrow/broad [NII] λ6583/Hα flux ratio of ~ 0.6. In the following we will treat the incidence of the firm detections as a conservative lower limit, but consider the average of this value and the incidence of firm and candidate detections (66±15%) as the most likely value of incidence.

As discussed in Section 3.1, the detectability of broad emission components does not vary much as a function of mass (indicated by the thick black and magenta curves in the right panel of Figure 3 and the upper right panel of Figure 9), such that a broad component of the same fraction should have been detectable throughout the stellar mass range spanned by our galaxies. Table 3 summarizes the incidence of broad components as a function of stellar mass.

We next studied the ***average*** profiles as a function of galaxy stellar mass, independently of whether individual profiles exhibit broad components or not, by weighted co-adding of the spectra of ***all*** SFGs in the nuclear regions and outer disk regions of the same galaxies, in each of the four mass bins. The resulting residual broad profiles, after removal of the narrow components in multi-Gaussian fitting (as described in 3.2.1.) are shown in Figure 7 for both the nuclear and outer disk regions (blue and grey lines, respectively). The extracted properties of these co-added profiles in the mass bins are summarized in Figure 9.

The nuclear and outer disk residual broad components of the co-added spectra in Figure 7 are basically identical in the lowest two mass bins (upper panels), and even in the third mass bin ($\log(M_*/M_\odot)$=10.6-10.9, lower left panel) the nuclear broad component on average is only marginally wider than its outer disk counterpart. Then in the highest mass bin (lower right panel) the broad nuclear component is drastically wider than the outer disk one. This suggests that on average the broad nuclear and outer disk components at $\log(M_*/M_\odot)$=9.4-10.9 reflect largely the same physical process, namely modest outflow velocity (~200 km/s) winds driven by massive stars and supernovae throughout the entire galaxy, as discussed in Genzel et al. (2011) and Newman et al.

(2012). Above log($M_*/M_\odot$) ~ 10.9 an entirely different physical process appears that originates only in the nuclear regions, and has much higher outflow velocities for almost all galaxies, thus completely changing the average nuclear spectrum. This is not to say that there are not a few such broad nuclear outflow sources at lower mass, but they are much rarer there, as seen from Figures 2 and 6.

### *3.3.2 Correlation of the broad nuclear components with bulge stellar mass*

Next we estimated the incidence of broad components as a function of bulge mass. This is motivated by the finding of several groups that the bulge mass (or central stellar surface density), and not the total stellar mass, appears to be most strongly correlated with the quenched (red) fraction at the high mass tail of the z=0-2.5 galaxy population (Franx et al. 2008, Cheung et al. 2012, Bell et al. 2012, Wake, van Dokkum and Franx 2012, Fang et al. 201, Lang et al. 2014).

Lang et al. (2014) have demonstrated that it is possible to infer high-z bulge masses from spatially resolved SED modeling of multi-band optical and near-IR HST imagery yielding stellar mass maps, and then carrying out a two-component structural analysis. In the right panel of Figure 6 we exploit the analysis of Lang et al. (2014) and Tacchella et al. (2014) for our SINS/zC-SINF and $KMOS^{3D}$ targets to explore the incidence of broad nuclear emission sources as a function of bulge mass. The quoted mass corresponds to that of the bulge from the best-fit two-component disk + bulge model (Sersic profiles with index n = 1 and n = 4, respectively) to the two-dimensional stellar mass distribution of the galaxies. The trend seen as a function of bulge mass is broadly similar to that as a function of total galaxy stellar mass in that there is a steep onset in the fraction of nuclear broad emission line galaxies, which occurs at/above $\log(M_{*,bulge}/M_{\odot})=10$.

However, we cannot distinguish on the basis of these comparisons whether stellar mass or bulge mass (or another quantity correlated with these, such as central black hole mass) is a better predictor of the onset of a nuclear broad component, presumably because of the combination of the uncertainties in the derived bulge masses, as well as the still modest size of our sample in view of the significant scatter in inferred bulge masses at a given galaxy's stellar mass.

***3.3.3 Properties of the broad nuclear components as a function of redshift***

Our sample is sufficiently large that we can compare the properties of the broad nuclear components in two different redshift bins, z=0.9-1.6 and z=2-2.6. Based on the results in the last sections we selected the 8, respectively 26, SFGs with $\log(M_{*}/M_{\odot})\geq10.9$ in these two redshift bins with firm or candidate broad emission components (implying incidences of 67 (±24) % and 81 (±16) %, respectively) and computed their SNR weighted co-added spectra. These are shown in the top panels of Figure 8. Qualitatively, emission line profiles with similar large [NII] λ6583/Hα ratios in their narrow and broad components and comparable broad to narrow flux ratios (upper right panel of Figure 9) are clearly detected in both redshift ranges, suggesting that the broad nuclear component phenomenon is present throughout the entire time period across the peak of the cosmic star formation epoch. However, the width of the z~1 broad component is only 800 km/s (in the Hα and [NII] lines), about half of that of the z~2 co-added profile (left panel of Figure 9). It is probably premature to assign a high significance to this tantalizing difference, given the smaller sample size at the lower redshift and the intrinsic large scatter of the broad line widths of the individual detections in both redshift ranges (left panel of Figure 9). Clearly a further improvement on the statistics in the lower redshift range is highly desirable.

### *3.3.4 Properties of the broad nuclear components as a function of specific star formation rate*

The left panel of Figure 6 shows the distribution of nuclear broad detections and candidates as a function of specific star formation rate. Within the statistical uncertainties the incidence of broad nuclear components (with or without candidates) does not seem to depend much on the vertical position in the stellar mass – star formation rate plane. The broad nuclear component profiles, broad to narrow and [NII] λ6583/Hα flux ratios above and below the main sequence also are qualitatively similar (bottom panels of Figures 8 and 9). In our decomposition of Figure 8, the average width of the broad component below the main sequence is twice as large as that above the main sequence. As with the similar difference between the average z=1 and z=2 profiles, this difference is tantalizing but it is not clear how much significance one should attach to it, given the large scatter in the individual line widths and the more modest sample size above the main sequence than below the main sequence.

Most surprisingly perhaps, we detect broad nuclear components just as likely significantly below the main sequence as we do near the main sequence, at least for those SFGs in which Hα is detected at all. The average width of the broad component in the co-added spectrum of the 4 SFGs (with firm and candidate detections) that lie much below the main sequence is as large as that for galaxies near the main sequence (bottom right panel of Figure 8).

The fact that the properties of the broad component depend little on specific star formation rate is highly interesting and informative in terms of the underlying physics. Tacconi et al. (2013), Magdis et al. (2012) and Saintonge et al. (2012) have presented evidence from molecular and dust observations that near the main-sequence sSFR correlates most strongly with galaxy baryonic gas fraction and star formation efficiency (the inverse of the gas depletion time scale). This suggests that the presence of the broad nuclear emission component is not strongly correlated with the gas properties on a galaxy wide scale. The fact that the broad nuclear emission component is also not more prominent for the few outlier SFGs in our sample (at sSFR/sSFR(ms,z)>4), including the very compact and high H$\alpha$ surface brightness source SA12- 6339, suggests that the nuclear broad line emission is also not primarily related to compact nuclear starbursts (see also section 4.3). Finally the detection of a broad component in galaxies one to two order of magnitude below the main sequence, in one of them with AO-assisted data (SDSS1030-2026), is very exciting indeed, as this shows that the same mechanism is likely operational in red-sequence galaxies.

# 4. Discussion

## 4.1 Mass outflow rates

In the following we estimate the mass outflow rates, as well as the momentum and kinetic energy transported in these (circum)-nuclear outflows. We assume that the nuclear broad emission represents an outflow into a cone of solid angle *Ω*, with a radially constant mass loss rate $\dot{M}_{out}$ and outflow velocity $v_{out}$. These assumptions are motivated by recent observations of the dependence of MgII absorber occurrence and profiles as a function of inclination of the host galaxy (Bordoloi et al. 2011, Kacprzak et al. 2011, 2012, Bouché et al. 2012), as well as theoretical work on both energy and momentum driven outflows (Veilleux et al. 2005, Murray et al. 2005, Hopkins, Quataert & Murray 2012). Following Veilleux et al. (2005) and Rupke et al. (2005) we take the wind outflow velocity to be the blue-shifted velocity at the HWHM of the broad profile, $v_{out} \sim |<v>_{broad} - 0.5 \times \Delta v_{broad}(FWHM)|$, which is a fairly conservative estimate of the intrinsic outflow velocity (see discussion in Genzel et al. 2011). We assume that the gas is photoionized, and in case B recombination with an electron temperature of $T_e=10^4$ K (Osterbrock 1989). In our simple model (c.f. Genzel et al. 2011) the ***average*** electron density and ***volume filling*** factor of the outflowing ionized gas scale with radius as $R^{-2}$ (for a constant mass outflow rate) but the *local* electron density of filaments or compact clouds from which the Hα emission originates does not vary significantly with radius and takes on a value of $< n_e^2 >^{1/2} \sim 80$ cm$^{-3}$. This choice is motivated by the average value of electron densities in the star forming ionized gas in the disks and centers of the SFGs of our sample, as derived from the [SII] λ6716/λ6731 ratio (<F(6716)/F(6731)> = 1.2 ± 0.06), combined with the assumption that the ionized gas in the outflows is in pressure equilibrium with that star forming gas as a result of shock excitation (section 3.2.2). Any departure from this assumption most plausibly drives electron densities in the outflows toward lower values, in which case the values for outflow rates and mass loading factor estimated below are lower limits.

For purely photoionized gas of electron temperature $T_4 = T_e / 10^4 K$ and case B recombination, the effective volume emissivity is $\gamma_{H\alpha}(T) = 3.56 \times 10^{-25} T_4^{-0.91}$ erg cm$^{-3}$s$^{-1}$, (Osterbrock 1989). The total ionized gas mass outflow rate, independent of Ω, can then be obtained from the extinction corrected, optically thin Hα luminosity $L_{H\alpha,0}$ via

$$
\begin{aligned}
&L_{H\alpha,0} = \gamma_{H\alpha}(T) \int \Omega R^2 n_e(R) n_p(R) dR, \\
&M_{HII,He} = \mu \cdot \int \Omega R^2 n_p dR = \frac{\mu L_{H\alpha,0}}{\gamma_{H\alpha}(T) \times n_e}, \text{ and} \\
&\dot{M}_{out} = \Omega R^2 \mu \, n_p(R) v_{out} = M_{HII,He} \cdot \frac{v_{out}}{R_{out}} \qquad (1).
\end{aligned}
$$

Here, $n_p$ is the proton density, $\mu = 1.36 \cdot m_p$ is the effective mass, for a 10% helium fraction, and $M_{HII,He}$ is the mass in ionized H and in He. $R_{out}$ is the outer radius of the outflow that initially is launched near the nucleus. We take $R_{out}$ as the half width at half maximum radius of the broad component emission, with $\langle R_{HWHM} \rangle$ ~1.25 kpc from an average of the spatially resolved data in FS14a and E. Wuyts et al. (2014b, in preparation).

To compute the intrinsic Hα luminosity for the broad component we corrected the observed fluxes for extinction using the visual extinction towards the bulk of stellar light $A_{V,\rm stars}$ from the best-fit SED models to the galaxies' SEDs (Section 2.1) and accounting for extra attenuation towards the nebular gas following the recipe $A_{V,gas}=A_{V,stars}\times(1.9 - 0.15\times A_{V,stars})$ found by S.Wuyts et al. (2013) as a best fit for the spatially resolved rest-UV to optical SEDs and Hα data of z=0.5-1.5 SFGs from the 3D-HST survey (see also Price et al. 2014). As for the SED modeling, the Calzetti et al. (2000) law was assumed to calculate the continuum extinction at the wavelength of Hα. This provides almost certainly a conservative lower limit to the intrinsic luminosity since the 80-140 km/s blue-shift of the broad line profile (relative to the narrow Hα emission) in several SFGs suggests a significant amount of differential extinction within the outflowing component (c.f. Genzel et al. 2011). The intrinsic Hα luminosity from the narrow component emission was computed in the same manner and used to derive the SFRs in the nuclear regions via the Kennicutt (1998) conversion adjusted to our adopted Chabrier (2003) IMF.

Table 4 summarizes the inferred mass outflow rates, the mass loading factors referred to the SFR in the nuclear regions, the ratios of outflow momentum rates to radiation momentum rates L/c, and the ratios of outflow kinetic energies to the luminosities of the (circum)-nuclear regions, for all 20 $\log M_* > 10.8$ SFGs with a good parameter definition of a nuclear broad component (excluding the three BLR sources). Figure 10 shows the resulting distributions of the inferred mass outflow rates and mass loading factors in histogram form.

Keeping in mind the large uncertainties of all the numbers, resulting in systematic uncertainties of the outflow, momentum and energy rates by at least a factor of 2 up and down, the median mass loading factors of the ionized outflows relative to the nuclear star formation rates are plausibly near/above unity, and the median outflow rates are about 100 $M_\odot$ $yr^{-1}$, comparable to the values of stellar feedback driven winds in the disks of these high-z galaxies (Erb et al. 2006, Genzel et al. 2011, Newman et al. 2012). Any additional contribution from very hot ionized plasma as well as cold atomic and molecular material in the outflows would increase this estimate.

The main physical difference between the nuclear-AGN and the disk-stellar feedback cases are the ***large outflow velocities*** (see left panel of Figure 9), not the mass loading and outflow rates (top right panel of Figure 9). The median outflow velocity of the nuclear outflows is ~500 km/s, more than twice that of the stellar feedback driven winds as estimated from the broad component in the outer disks and at lower masses ($v_{out}$(disk)~200 km/s). In 6 cases the nuclear outflow velocity exceeds 700 km/s. Higher outflow velocities for AGN feedback is also characteristic for gas-rich, luminous AGN-ULIRGs at low-z (Sturm et al. 2011, Veilleux et al. 2013, Rupke & Veilleux 2013, Spoon et al. 2013). This means that in about half of the nuclear outflow galaxies in Table 4 the outflow velocity is at least twice the rotation velocity of the galaxy, implying that the nuclear outflows in principle can fully escape the galaxies, and perhaps even their halos. That is obviously not the case for the disk outflows. The stellar feedback likely only drives fountains where the gas will return after about a billion years or less, as indicated by recent theoretical work (Davé, Oppenheimer & Finlator 2011, Zhang & Thompson 2012, Übler et al. 2014).

The median ratio of the momentum in the outflows to that in (stellar) radiation is ~5, and there are 9 SFGs where this value is 10 or more. Such large values probably argue against momentum driven outflows (Dekel & Krumholz 2013, Krumholz & Thompson 2013). The median energy in the outflows is ~0.4 % of the nuclear star formation luminosities. Theoretical estimates suggest that energy driven outflows can account for up to 1 % of the energy source (Murray, Quataert & Thompson 2005). Taking the nuclear star formation luminosities estimated from the narrow Hα emission as a guide, radiation energy driven outflow would be possible for one half but not the other half of the sample in Table 4.

For those of our SFGs with AGN identifications (see section 4.2), we have used the absorption corrected X-ray luminosity, and/or the mid-IR luminosity, or a combination of both, to estimate the bolometric AGN luminosity, using the techniques of Rosario et al. (2012, for X-rays) and Richards et al. (2006, for mid-IR). Despite inhomogeneous data, it is natural to assume that these identified AGN preferentially sample larger AGN luminosities among our targets. We list these luminosities in the next to last column of Table 1. If we only had a mid-IR estimate we assumed that this constitutes effectively an upper limit to the AGN luminosity because of contributions to the mid-IR luminosity by dusty star formation. The last column of Table 1 gives the ratio of the galaxy integrated luminosity from star formation to this AGN luminosity estimate. That ratio varies over more than an order of magnitude from source to source but on average has a value of 1.3. Since the nuclear star formation rates typically are 30-40% of the galaxy integrated star formation rates, the mass loading factors, as well as momentum and energy ratios in Table 4 would decrease by a factor of ~2, when compared to the AGN luminosity, rather than to the nuclear stellar luminosity. This may increase the probability that the nuclear outflows are momentum driven if the AGN is active, although this conclusion carries substantial uncertainty.

Mechanical driving of the nuclear outflows may be an additional possibility, as in other low-luminosity AGNs and black hole systems (Fabian 2012, McNamara & Nulsen 2007).

Massive high-z SFGs near the main-sequence are gas rich, with typically $>10^{10}$ $M_\odot$ of molecular gas in the central few kpc (Tacconi et al. 2013). Our observations imply that these circum-nuclear gas reservoirs can in principle be driven out by the nuclear outflows over a time scale of a few hundred Myrs. If there is efficient radial transport of gas from the outer disk to the center, as advocated by many theoretical studies (Noguchi 1999, Immeli et al. 2004 a,b, Genzel et al. 2008, Bournaud, Elmegreen & Martig 2009, Ceverino, Dekel & Bournaud 2010, Dekel & Burkert 2014, Forbes et al. 2014), the nuclear outflows may even be an efficient process for removing gas from the entire galaxy.

## 4.2 Correlation with X-ray/optical/infrared/radio AGN

### *4.2.1. Identification of AGN*

In this section we analyze the relationship and relative incidence of the nuclear broad emission SFGs discussed in the last section, to the AGN populations in the same cosmological fields.

For this purpose we searched for signatures of contemporaneous nuclear activity in our sample of massive galaxies using five different tracers. X-ray imaging and catalogs are available for 91 of the 110 galaxies in Table 1 from the Chandra Deep Fields North/South (Alexander et al. 2003, Xue et al. 2011, Brightman & Ueda 2012), Extended Chandra Deep Field South (Lehmer et al. 2005), AEGIS-X survey (Laird et al. 2009),

Subaru XMM Deep Field (Ueda et al. 2008) and the SDSS J1030+0524 QSO field (Farrah et al. 2004). These fields vary considerably in instrumental coverage and depth, from 4 Msec with Chandra in the CDF-S to 86 ksec with XMM-Newton in the SDSS1030 pointing, spanning sensitivities going down to X-ray emitting star-forming galaxies in the deepest data to fairly luminous AGN with X-ray luminosities of $>10^{44}$ erg/s (z~2) in the shallowest field. Nevertheless, we proceed knowing that we may be missing a proportion of active galaxies from our sample. In total, 13 of the SFGs in Table 1 are detected in the X-rays, of which 11 are confirmed AGN based on various X-ray diagnostics as developed by Xue et al. (2011).

Spitzer imaging and public catalogs in the four IRAC bands are available for 91 galaxies, from the GOODS-S survey (Dickinson et al. 2003), SWIRE survey (Lonsdale et al. 2003), AEGIS survey (Barmby et al. 2008) and SCOSMOS (Sanders et al. 2007). While the depths of the IRAC data do vary between fields, the coverage is more uniform than among the X-ray datasets. We use the criteria of Donley et al. (2012), which identify AGN based on their observed IRAC 5.8μm/3.6μm to 8.0μm/4.5μm flux ratios. This method is fairly free of contamination from starbursts at z~2, but may miss some weak AGN. The IRAC flux ratios of our sample SFGs are plotted in Figure 11, along with contours indicating for reference the distribution of IRAC-detected objects from S-COSMOS in the same range of z ~ 0.7 – 2.6, and red lines enclosing the AGN selection wedge according to Donley et al. (2012). In total, 9 galaxies satisfy the IRAC AGN criteria, with 15 more potential AGN that lie close to the region delineated in Donley et al. (2012) but formally do not satisfy the criteria.

Some of the fields from which the samples are drawn have VLA 20 cm radio catalogs from VLA-COSMOS (Schinnerer et al. 2010), AEGIS-20 (Ivison et al. 2007), GOODS-N (Morrison et al. 2010), ECDF-S (Miller et al. 2013) and SXDF (Simpson et al. 2006). Of the 75 galaxies in our sample with radio coverage, 7 are detected at the depths of the corresponding surveys. Since all these fields are also covered by Spitzer MIPS imaging and catalogs, we used the 24μm to 20 cm observed flux ratio as a way to discriminate between true radio-loud AGN and galaxies dominated by star-formation in the radio band, following the approach of Appleton et al. (2004), but including a k-correction based on the typical star-forming galaxy SED from Wuyts et al. (2008). Only 1 galaxy ($KMOS^{3D}$-GS3-18419) is identified as radio-loud and its AGN nature is also confirmed by IRAC-based criteria.

We also used available rest-frame UV spectroscopy to search for the standard AGN emission line indicators. Four of the SFGs in Table 1 are identified as AGN in that way (BX663, D3a15504, J0901+1814 and $KMOS^{3D}$-GS3-19791), as has been previously pointed out by Förster Schreiber et al. (2011, 2014a) and Fadely et al. (2010).

In addition to the methods described above, which apply to a large fraction of the galaxies, we also searched published samples of AGN selected by variability in the optical or X-ray in the GOODS and ECDFS fields (Trevese et al. 2008, Villforth et al. 2010, Young et al. 2012) and samples of galaxies searched for VLBI radio cores (Middelberg et al. 2011, Chi et al. 2013). Only GOODSN-22747 was identified as an AGN in these studies, consistent with its independent identification as an X-ray AGN.

In total, we have X-ray, mid-IR, radio or optical spectroscopy data relevant to AGN identification on 95 of the 110 SFGs in Table 1 (henceforth called the ‘common sample’).

### *4.2.2. AGN incidence as a function of stellar mass*

With the AGN identifications and candidates from the last section, we find that the AGN incidence strongly varies with galaxy stellar mass, qualitatively mirroring the incidence of the broad nuclear components discussed in section 3.3.1. Figure 12 compares the broad component with the AGN fractions in the common sample (with both AGN and broad component data) as a function of stellar mass for our SFGs, and with the fraction of AGN among the more general population of z ~ 1 – 2 SFGs. The green/brown and yellow/green asterisks in Figure 12 denote the AGN fractions in the common sample for the 'firmly identified' AGNs and the firm plus 'candidate' AGNs. The AGN fraction for $\log(M_*/M_\odot) \geq 10.9$ in the 'common sample' is 38 (±10) %. Including the AGN candidates the value would increase this value to 51 (±12) % (Table 3). The grey and green shaded distributions in Figure 12 denote the AGN incidence expanded to the entire GOODS N/S and COSMOS fields but corrected upward by 30% to estimate the AGN fraction in the star forming population only. According to this estimate the AGN incidence at $\log(M_*/M_\odot) \geq 10.9$ is 28 (±10)%. All these values are in good agreement with previous findings in the literature, although statistical uncertainties are obviously large (e.g., Kauffmann et al. 2003b, Reddy et al. 2005, Papovich et al. 2006, Daddi et al. 2007, Brusa et al. 2009, Xue et al. 2010, Hainline et al. 2012, Bongiorno et al. 2012, Rosario et al. 2013b).

At face value the incidence of broad components at $\log(M_*/M_\odot) \geq 10.9$ is about 1.5 times larger than those of the AGN in the common sample. If the estimates of the broader COSMOS and GOODS fields are used, that ratio increases to between 1.8 and 3.5.

Overall the data thus may suggest that the nuclear broad emission activity has approximately twice the duty cycle of AGNs in this highest stellar mass bin at/above the Schechter mass. Caution is warranted, however, to not over-interpret this potentially very interesting difference. The statistical uncertainties alone are already large enough to make up some of the difference in incidence. If one takes 0.66 as the best estimate of the broad component fraction at $\log(M_*/M_\odot)\geq 10.9$ (the average of the firm nuclear outflow sources and the number including candidates), and 0.37 as the AGN incidence (an average of the firm AGN and firm plus candidates in the common sample, and the COSMOS and GOODS numbers), the difference is statistically significant at the ~2.5σ level. In addition the aforementioned variations in depth of the AGN indicators in the different fields, along with the possible effects of extinction and AGN variability would systematically increase the AGN fraction and thus further decrease the differences.

We take a conservative approach and conclude from the current evidence that the strongly mass-dependent ***incidence of broad nuclear components is at least as large as that of AGN***. However, if the identification of many of our candidates as broad line sources were to be confirmed, and/or statistical uncertainties further reduced, it is possible that the incidence of nuclear outflows exceeds that of luminous AGNs by a factor ~2. Because of the more homogenous coverage and lower susceptibility to variability and extinction, the occurrence of the broad emission may likely turn out to be a better way of characterizing the impact of massive nuclear black holes on their surroundings than the AGN light/activity.

The issue of AGN variability in particular has recently been pointed out as a major stumbling block in investigating the co-evolution of massive black holes and their host galaxies (e.g. Hickox et al. 2014). High-z AGN of luminosity as detected in X-ray deep fields vary by few tenths of a magnitude over a few year time scale (Salvato et al. 2011, Wold et al. 2007). Studies of local AGN suggest a power law slope -1 in the power spectral density of light curves ($P(\nu) \propto \nu^{-\alpha}$ with $\alpha \approx 1$, e.g., McHardy et al. 2006, Webb & Malkan 2000). While it is difficult to extrapolate to very low frequency $\nu$, large variations have been observed in the few luminous AGN that were monitored over decades (Ulrich et al. 1997). Very large variations of the AGN may occur over the response time of a kpc-size photoionized outflow region, which will be at least thousands of years due to combined light travel and recombination timescales (where recombination time may be shorter, depending on local electron density). Observability of the outflow may be extended further into periods of unobservable direct AGN radiation if the ionizing agent is a combination of photoionization and of delayed shocks set by the AGN outbursts, as discussed in section 3.2 (e.g., Zubovas & King 2012; Gabor & Bournaud 2014). The dynamical crossing time of the nuclear outflow regions is about 3 million years, smoothing out any variability in the nucleus and making the outflow still observable when the AGN is off or weaker. The recombination time scale in the winds and nuclear narrow line gas is probably less than the light travel time, calling for an ionization agent when AGN radiation levels are low (FS14a).

## 4.3 Can nuclear star formation bursts drive the nuclear outflows?

We have shown in section 3.2.2. and in Figure 5 that for a fraction of the broad nuclear outflow sources their narrow line ratios cannot be explained by stellar photo-ionization, but require an AGN or a combination of shocks, AGN and stellar ionization.

Based on the very high [NII] λ6583/Hα ratio in the stacked broad nuclear component of all $\log(M_*/M_\odot)\geq10.9$ SFGs in the lower right panel of Figure 7, this conclusion can plausibly be extended to the average galaxy in this mass range. We have also shown that the incidence of a distinct nuclear outflow component (of much greater inferred outflow velocity than in the extended 'disk' outflows) increases rapidly at or above the Schechter mass (section 3.3). And finally, we have shown in the last section (4.2) that the incidence of AGN as identified in X-ray/mid-IR/radio/optical spectroscopy tracers also increases rapidly at and above this mass. Taken together, these findings provide strong circumstantial evidence, but by no means a unique proof that the broad nuclear outflows at $\log(M_*/M_\odot)\geq10.9$ are driven by the central massive black holes.

Another constraint of the relative roles of (circum-) nuclear star formation and AGN in accounting for the broad (circum-) nuclear outflows comes from the nuclear concentration of the narrow Hα emission, which should track star formation. Under the counter-hypothesis that (circum-) nuclear star formation, and not AGN, is the main driver also of the nuclear outflows (as well as the disk outflows), one would then expect that the concentration of narrow Hα emission is more pronounced in those SFGs with well detected central outflow components, than in the SFGs without individual detections. Enhanced extinction in the nuclear regions would weaken such central peaks in narrow Hα emission, but at the same time plausibly also the broad emission.

We have measured the ratio of narrow Hα emission in the circum-nuclear region (0.6" diameter for seeing limited, and 0.35" for AO data, as described in section 2.2) to the galaxy integrated narrow Hα flux for all 52 SFGs with IFU data at $\log(M_*/M_\odot) >10.5$ for which this ratio could be reliably determined. In the remaining 11 SFGs with IFU data the narrow emission is either too faint, or the Hα emission is totally dominated by broad emission. Of these 52 SFGs 28 have individually detected nuclear broad components (of quality 0.5, 1 or 2), 24 do not. The medians/means and standard deviations of the ratio of nuclear to total narrow Hα flux are 0.19 (±0.1) and 0.18 (±0.09) for the SFGs with and without individually detected nuclear broad components, respectively. The resulting uncertainty of the mean in both groups is ±0.02. For comparison the flux ratio for a point source is 0.6 (±0.05), such that the narrow Hα emission in 50 of the 52 SFGs is significantly extended in our data. The distributions and centroids of the ratio of nuclear to total narrow Hα flux for the two groups of SFGs are thus statistically indistinguishable, and the hypothesis that nuclear star bursts solely account for the (circum-) nuclear outflows can be rejected.

We thus conclude that the nuclear outflows are likely driven by the central massive black holes.

## 4.4 Stellar mass estimates for AGN hosts

Given the common presence of AGN among our sample of galaxies as discussed in the last sections, a potential concern is the reliability of the stellar masses we have been using because the emission from the AGN itself can contribute significantly or even outshine the rest-UV to near-IR emission from the stellar populations of the host galaxy. In the worst case, the inference of a mass threshold might be largely driven by the presence of a luminous AGN artificially driving up the inferred stellar masses.

From the comparison of SED fitting based on stellar population synthesis models (as used to derive the $M_*$ and SFR estimates of our SFGs; see Section 2.1) versus a more detailed decomposition accounting for both stellar and AGN light, Santini et al. (2012) showed that stellar masses for type 2 AGN are typically well recovered with pure stellar templates (with differences on average consistent with zero, a scatter within a factor of two, and ~ 1% of objects having larger differences). In contrast, type 1 AGN, whose SEDs are generally more dominated by AGN light, were found to exhibit a much larger scatter of a factor of ~ 6, with ~ 30% of the objects having stellar mass estimates differing by more than a factor of two, although the distribution was broadly consistent with typical differences of zero. These results are attributed to the significantly different AGN contributions to the observed SEDs between the two types. This behavior is also seen in the SEDs of our galaxies, plotted in Figure 13. Except for the three candidate BLR sources identified by their very broad line widths and their lack or weakness of forbidden line and narrow star-formation dominated emission (see Section 2.1 and Table 1), the SEDs of all galaxies including those with broad nuclear outflow signatures are consistent with being dominated by stellar emission: all show a strong Balmer/4000Å break. The three BLR candidates show instead very blue and fairly featureless SEDs.

We conclude from this inspection that the stellar masses are very likely sufficiently reliable for most of the AGN and broad nuclear outflow galaxies among our sample, and that possible associated uncertainties would not significantly affect our main findings. The stellar mass estimates for the three BLR candidates are more uncertain.

## 4.5 Connection to recently proposed progenitor candidates of compact quiescent galaxies

Barro et al. (2013, 2014) have pointed out the presence of a population of compact z~2 SFGs with large mass surface densities and velocity dispersions (see also Nelson et al. 2014), which may be candidate progenitors of the compact quenched galaxy population in this redshift range (e.g., van Dokkum et al. 2009; Bezanson et al. 2013; van de Sande et al. 2013; Belli et al. 2014). It is interesting to note that of the 31 $\log(M_*/M_\odot)\geq10.9$ SFGs with firm and candidate broad nuclear components, 18 (58%) fulfill the criterion $\log(M_*/R_e^{3/2})\geq10.4$, and 24 (77%) fulfill a slightly more relaxed criterion $\log(M_*/R_e^{3/2})\geq10.0$, as proposed by Barro et al. to identify candidate progenitors of high redshift compact quiescent galaxies. The overlap between the Barro et al. high surface density SFGs and our nuclear outflow galaxies is substantial. Future work needs to explore in more detail the relation between such compact SFGs and the broad nuclear outflow phenomenon.

# 5. **Conclusions**

From high quality seeing limited and adaptive optics observations with the SINFONI, KMOS, GNIRS and LUCI near-infrared spectrometers, we have extracted nuclear (radius < 2.5 kpc) and outer disk Hα/[NII]/[SII] spectra for a sample of 110 z=0.8-2.6 'normal' star forming galaxies with a roughly homogeneous coverage in the stellar mass – specific star formation rate plane. Compared to our previous work (FS14a) we have increased by a factor of six the critical number of SFGs near and above the Schechter mass: 74 SFGs above $\log(M_*/M_\odot)=10.6$ and 44 SFGs above $\log(M_*/M_\odot)=10.9$.

We fully confirm the presence of a very common occurrence of a broad (circum)-nuclear component (FWHM~450 - 5300 km/s) whose incidence is strongly mass dependent and not present in the outer disk spectra, in excellent agreement with FS14a. Depending on the quality cut on the individual spectra, at least half and perhaps as much as 90% of the SFGs in the mass bin $10.9 \le \log(M_*/M_\odot) \le 11.7$ appear to show this component, while below that threshold the occurrence drops sharply. The broad nuclear component is present above and below the main sequence of SFGs , including in several cases more than an order of magnitude below (in specific star formation rate) the main sequence, and across redshift from z~0.8 to 2.6, with roughly comparable width and in approximately similar strength relative to the narrow Hα emission.

The broad component is present in Hα, [NII] and [SII]. It is spatially resolved in a subset of AO-assisted SINFONI data sets (FS 14a) and one massive lensed galaxy (E. Wuyts et al. 2014b, in preparation), with a diameter of 2-3 kpc. This demonstrates that the component cannot be bound and must represent a powerful ionized nuclear wind on the scale of the classical narrow-line region of AGN.

From the ratio of broad to narrow line fluxes in our sample, we estimate the mass loading of the warm ionized outflow component, $(dM_{out}/dt)/SFR$, to be near unity, for a local wind electron density of 80 $cm^{-3}$. If so the nuclear outflows may in principle be able to eject a significant fraction of the circum-nuclear gas out of the galaxy, and help in quenching star formation at the high mass end of the star forming population.

For a subset of SFGs in which [NII] λ6583/Hα, [SII] λλ 6716+6731/Hα, [OI] λ 6300/Hα and [OIII] λ 5007/Hβ are detected the line ratios suggest that the most likely ionization/excitation source of the nuclear outflow and nuclear narrow emission is an AGN. Alternatively a combination of shock excitation with stellar photoionization is also possible.

The ~66% incidence of broad nuclear emission components in the highest mass bin is at face value about twice larger than, but statistically perhaps just consistent with the incidence of AGNs in the GOODS/COSMOS fields (~30%) , from combined X-ray, optical, infrared and radio indicators. If this difference is real, it might be caused by AGN variability/duty cycle or extinction. Central massive black holes may drive variable or episodic outflow components that then are still observable when its radiation (the AGN) is in a low state.

Reports on outflows in AGN at all redshifts abound in the literature. Our findings thus might at first not appear surprising. However, ***the key difference is that we selected galaxies on the basis of stellar mass and star formation rate, and not on the (highly variable) AGN luminosity. Our results thus imply that the majority of all galaxies at the massive tail of the population exhibit powerful outflows***.

How much can the statistics be expected to improve in the next few years? Within the next year or two we hope to increase the $KMOS^{3D}$ sample at high masses by 50%, including a better coverage below the main sequence, and in the redshift range 1 to 1.5. Including other ongoing surveys with KMOS at the VLT and MOSFIRE at the Keck telescope, one probably can hope for an increase to about 100 galaxies in that mass range, thus opening an excellent opportunity of mapping out the parameter dependences in more detail.

***Acknowledgements.*** *We thank the ESO Paranal staff for their excellent support with the SINFONI and KMOS observations for this work. We thank the referee for constructive and useful comments, which helped to improve the manuscript. The LBT is an international collaboration among institutions in the United States, Italy, and Germany. LBT Corporation partners are: The University of Arizona on behalf of the Arizona university system; Istituto Nazionale di Astrofisica, Italy; LBT Beteiligungsgesellschaft, Germany, representing the Max-Planck Society, the Astrophysical Institute Potsdam, and Heidelberg University; The Ohio State University, and the Research Corporation, on behalf of The Universities of Notre Dame, Minnesota, and Virginia. DJW and MF acknowledge the support of the Deutsche Forschungsgemeinschaft via Project ID 387/1-1.*

# Figures

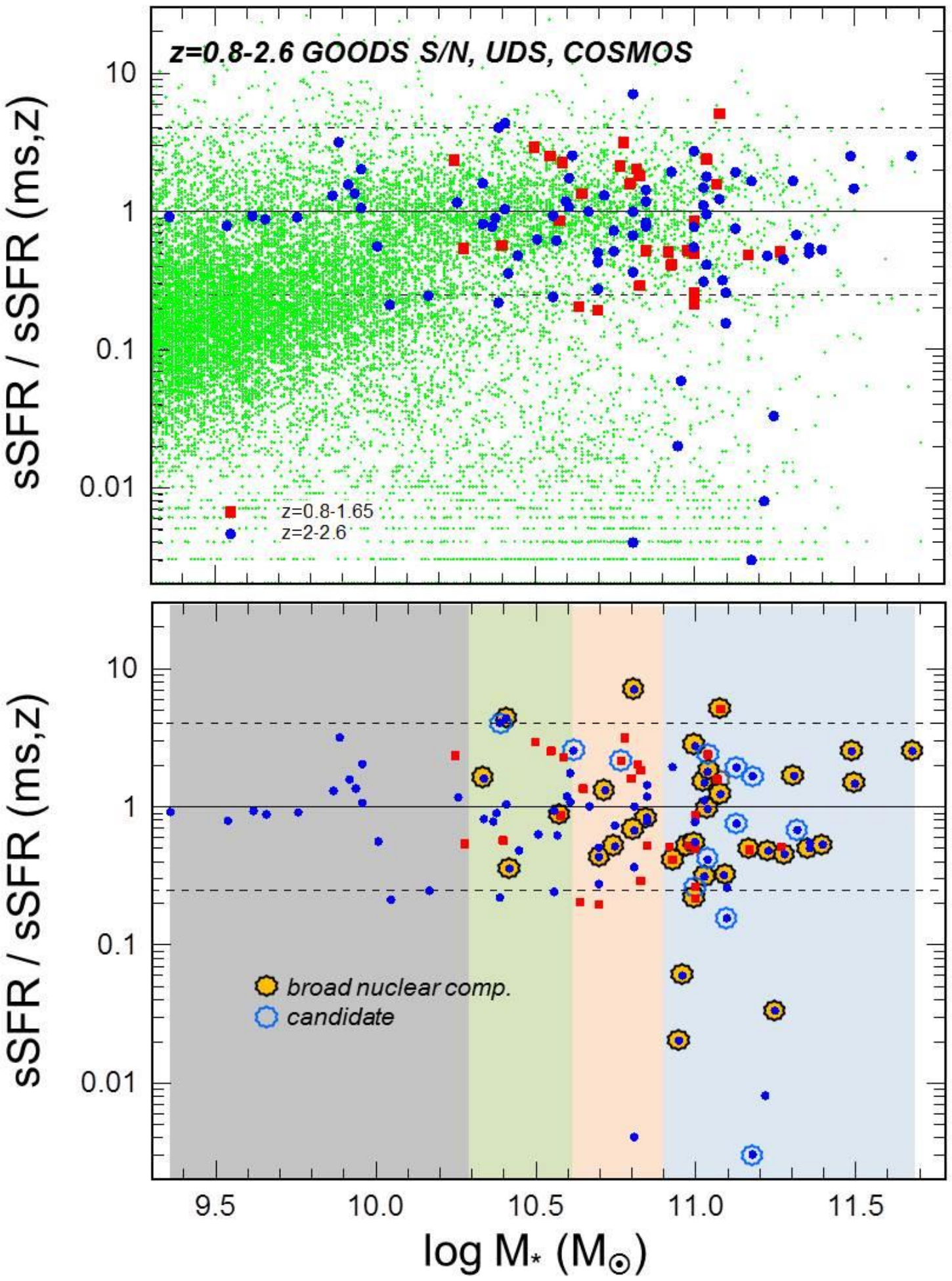

Figure 1. Top panel: Location of our final z=0.8-2.6 SFG sample of 110 galaxies in the stellar mass – specific star formation rate (sSFR) plane. We have divided the sSFR of each galaxy by the value of the main sequence line (as determined from the Whitaker et al. (2012) fitting function valid for $>10^{10}$ $M_{\odot}$) for a fair comparison of galaxies at different redshifts. Red squares denote z~0.8-1.6 and blue circles z~2-2.6 SFGs from the SINS/zC-SINF surveys (Förster Schreiber et al. 2009, 2014b, Mancini et al. 2011), the LUCI survey (E.Wuyts et al. 2014a), the first-year KMOS$^{3D}$ survey results (Wisnioski et al. 2014), and the GNIRS+SINFONI survey of massive galaxies by Kriek et al. (2007).

The small green dots represent the samples drawn from 3D-HST survey catalogs of z=0.8-2.6 galaxies in the CANDELS, GOODS N/S, COSMOS and UDS fields (e.g. Wuyts et al. 2011a,b, Brammer et al. 2012, Skelton et al. 2014). Bottom: The shaded vertical regions denote the four mass bins discussed throughout the text (grey: $\log M_*$=9.4-10.3, green: 10.3-10.6, pink: 10.6-10.9, blue: 10.9-11.7). Large orange filled black circles denote those galaxies in which the individual nuclear spectra exhibit a significant broad component. Open blue circles denote less certain candidates with possible broad components.

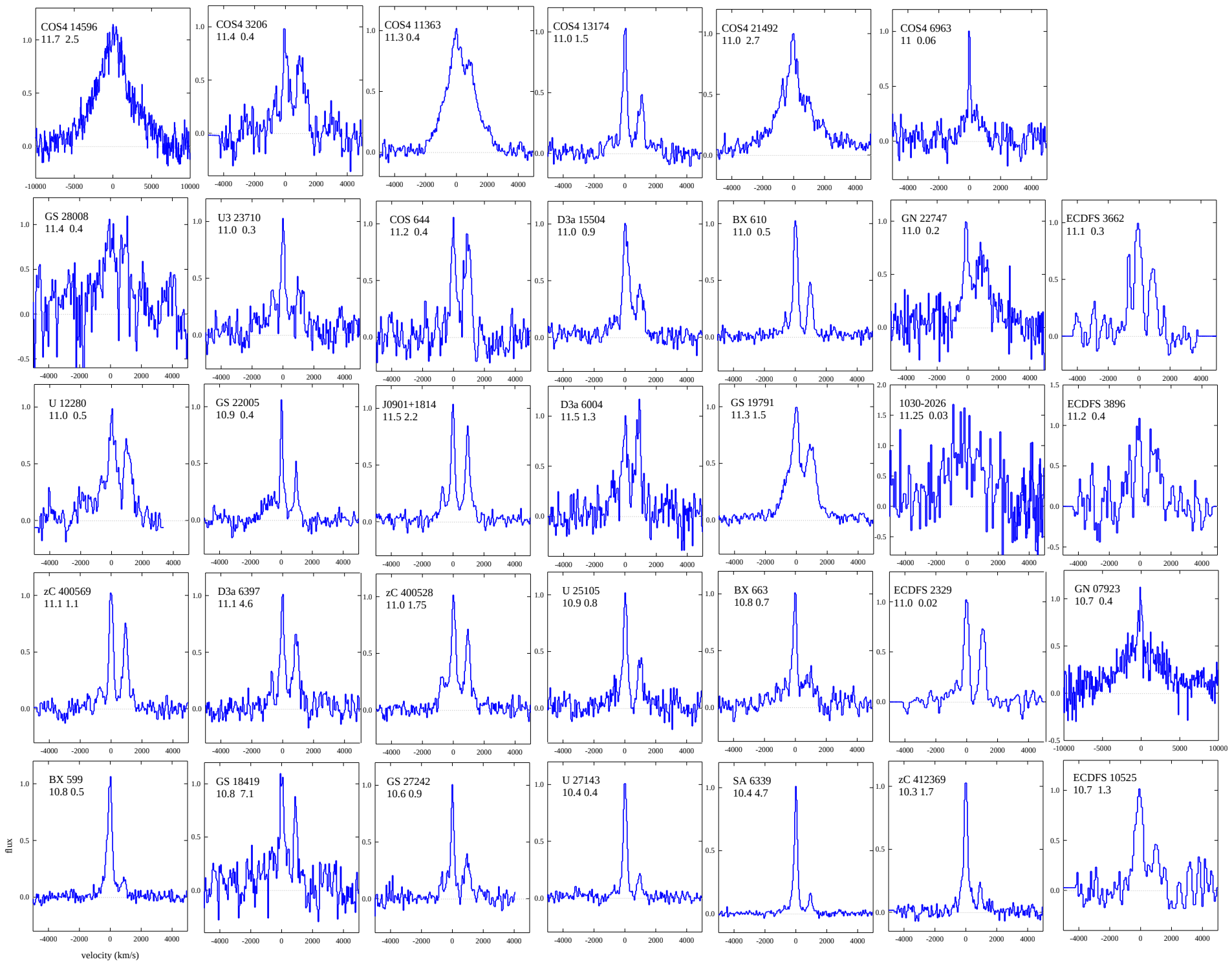

Figure 2. Individual nuclear spectra (extracted in apertures of FWHM 0.3-0.4” for AO data and 0.6” for seeing limited data) for the 34 SFGs with a firm detection of a broad component at the nucleus (quality ‘1’ or ‘2’ in column 7 of Table 1).

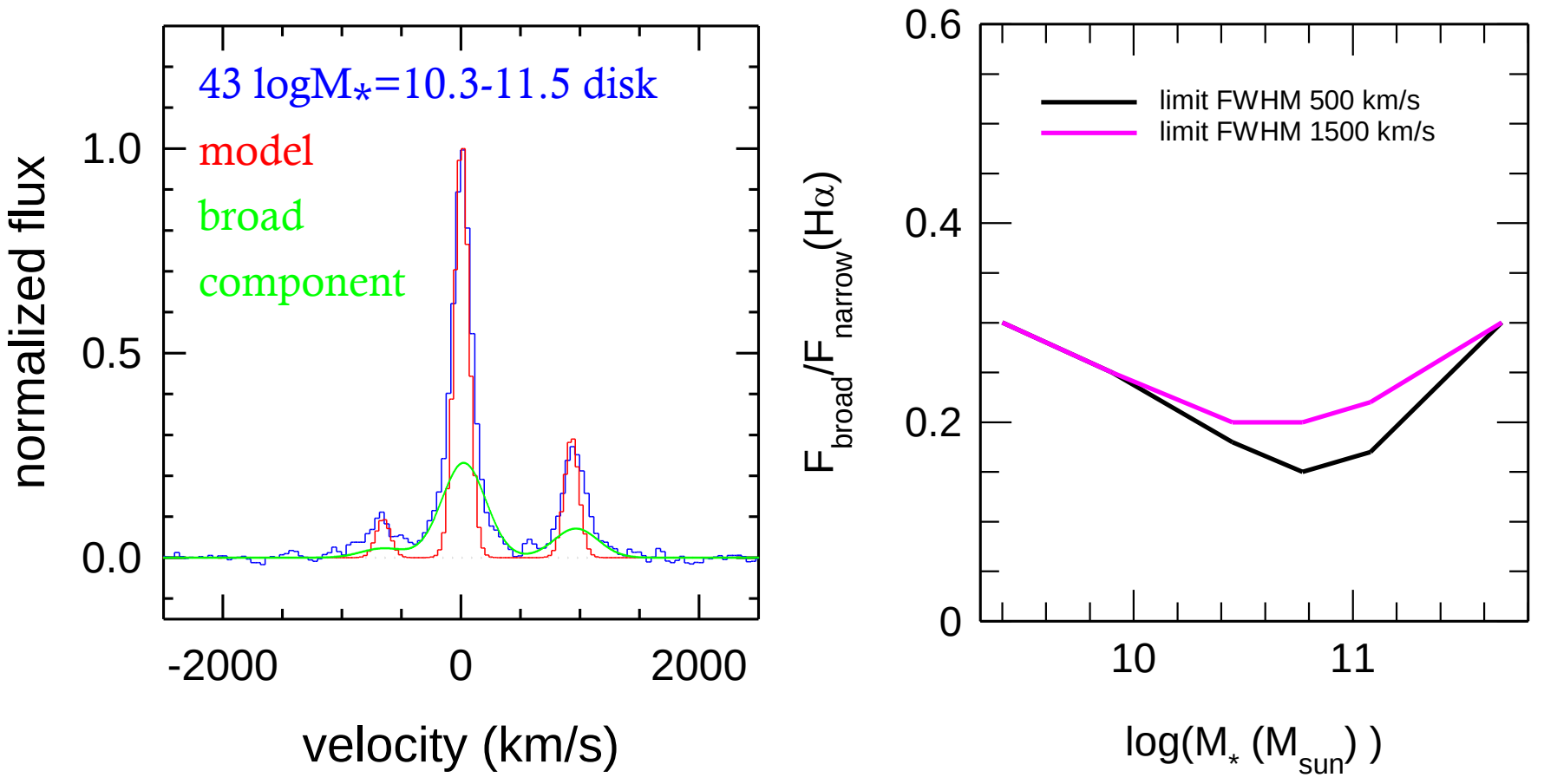

Figure 3:Left: Average outer disk spectrum of SFGs with log($M_*/M_\odot$)=10.3-11.5, from an equal weight co-add of 43 galaxies (blue), along with the best fit broad component in Hα+[NII] (green), which has FWHM~400 km/s. The red spectrum represents a massive model galaxy with a bulge and a disk ($M_{total}=1.5x10^{11}\ M_\odot$), resulting in a fairly flat intrinsic rotation curve of $v_{rot}$~240 km/s, observed at inclination $52^0$. The model data cube was convolved with a FWHM angular resolution of 0.55” and a FWHM spectral resolution of 100km/s, and then analysed in the same way as our SINFONI and KMOS data, removing the large scale velocity gradients from the rotation pixel-by-pixel and then extracting an outer disk spectrum at R>0.4”. The simulated spectrum has a FWHM ~160 km/s, but is clearly much narrower than the average disk spectrum of our sample. Since the model galaxy’s mass and rotation velocity is at the upper bound of our sample, the red spectrum indicates the maximum impact of residual beam-smeared rotation even in the seeing limited $KMOS^{3D}$and SINS/zC-SINF data sets. The broad emission in the disk spectrum thus must come from a gravitationally unbound component, as proposed earlier (Shapiro et al. 2009, Genzel et al. 2011, Newman et al. 2012). Right: Limits for detection (and correct parameter extraction) of broad components of FWHM 500 km/s (black) and 1500 km/s (pink), as a function of stellar mass, in the SFG stacked spectra analysed in this paper. These limits were derived by inserting Gaussian model components of different amplitudes and widths into the disk/nuclear co-added spectra (without significant broad components) and re-extracting their properties from 6 component Gaussian fits

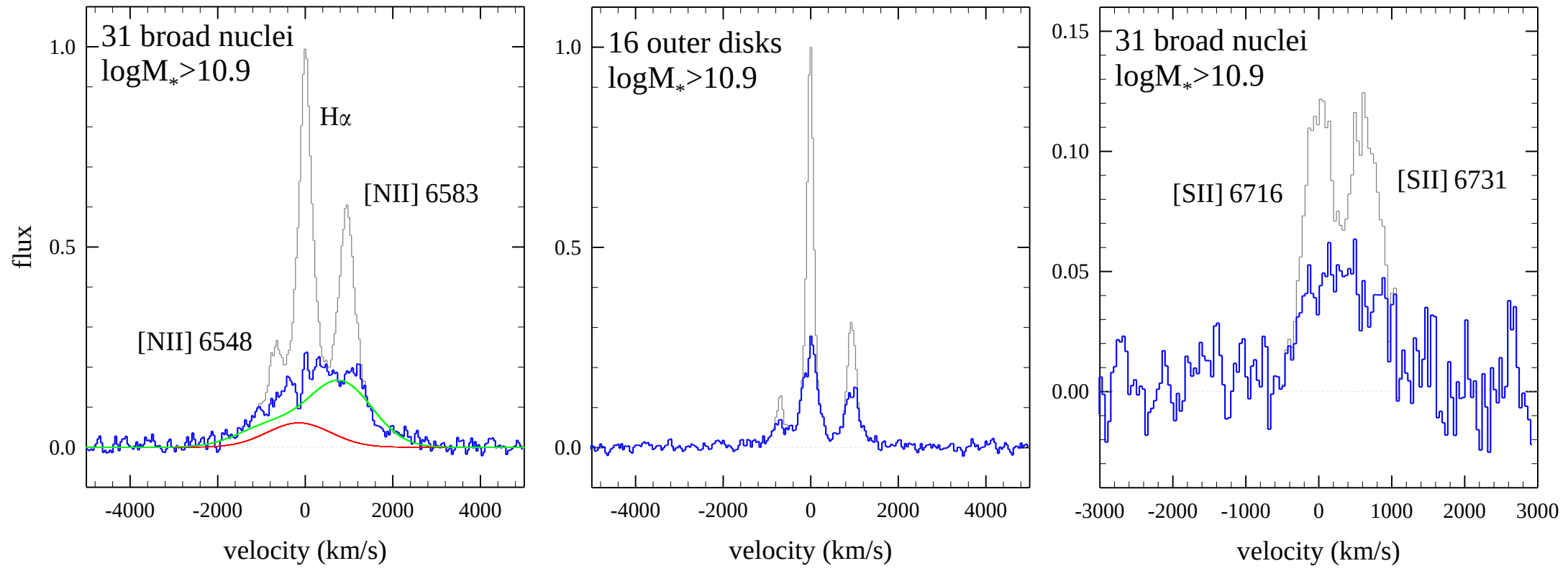

Figure 4. Left panel: Co-added Hα-[NII] spectrum (weighted by signal to noise ratio) of the 31 log($M_*/M_\odot$)=10.9-11.7 nuclei with individual, firm and candidate broad emission detections (grey), but excluding the 2 nuclei with broad line regions. The blue line denotes the broad component, after removal of the narrow Hα/[NII] lines, from a 6 parameter Gaussian fit. The thin dotted red and green curves show the Hα and [NII] broad fit components separately. Middle panel: Average outer disk spectrum (grey) for those 16 (of the 31) log($M_*/M_\odot$)≥10.9 SFGs for which significant extended Hα emission is detected, weighted again by SNR.  As in the left panel, the blue profile denotes the residual broad emission component. Right panel: Co-added [SII] spectrum (grey) of the 31 nuclei. As in the other panels, the blue profile denotes the [SII] broad component, after removal of the narrow [SII] λλ6716+6731 emission, assumed to have the same width and central velocity as the Hα/[NII] lines.

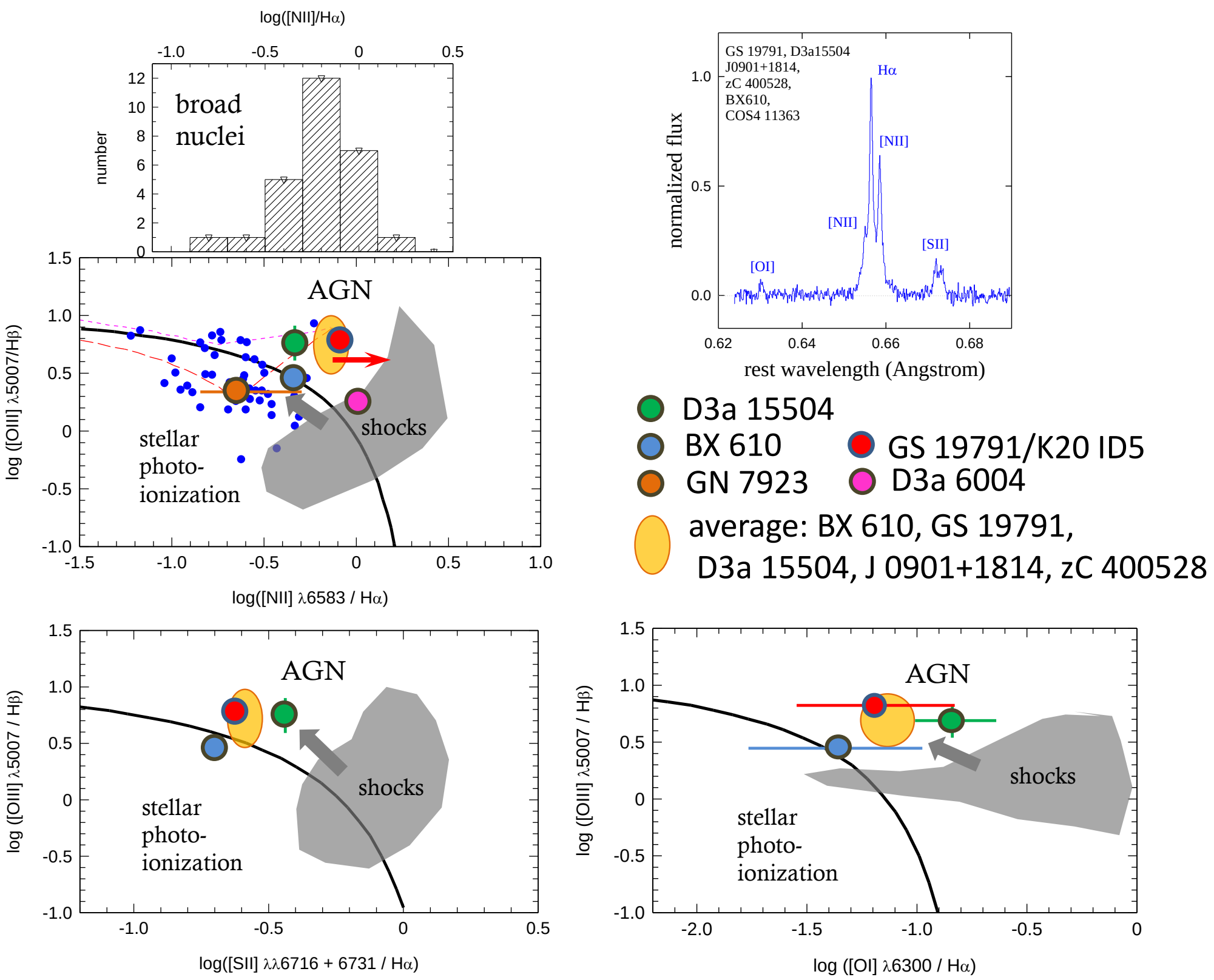

Figure 5. Diagnostic line ratio diagrams for the nuclear broad line SFGs in our sample. The three nuclei of GS3-19791, D3a-15504 and BX 610 have detections in all 4 ratios [NII] λ6583/Hα, [OIII] λ5007/Hβ, [SII] λλ 6716+6731 /Hα and [OI] λ 6300 /Hα and are plotted as large red, green and blue circles. D3a-6004 has two line ratios but the [OIII] λ5007/Hβ ratio refers to the galaxy as a whole. The large orange ellipse denotes the co-added spectrum of GS3-19791, D3a-15504, BX 610, J0901+1814 and zC400528 (top right panel). The red arrow pointing to the right indicates that for the broad line component, the [NII] λ6583/Hα ratio is a factor of about 2 larger than for the narrow component. Hatched black histograms denote the distribution of the (total) [NII] λ6583/Hα ratio in all SFGs of our sample that have a good detection of a nuclear broad

component (with the exception of zC400569, see text). The small filled blue circles are other z~1-2.5 SFGs from Newman et al. (2014), Trump et al. (2013), Shapley et al. (2005), Kriek et al. (2007) and Liu et al. (2008) (see also Steidel et al. 2014). The thick black line is the extremal 'starburst' line from the models of Kewley et al. (2001). Sources to the left of that line can be accounted for ISM photoionized by stars. The red-dashed line denotes the location of sources with a combination of a 'normal photoionized ISM' and the metal rich narrow line region around and AGN. The magenta-dotted line denotes the location of sources with a combination of an 'extreme photoionized ISM (large ionization parameter, high density)' and a metal rich narrow line region around an AGN (from Kewley et al. 2013). The large dark-grey polygons labelled 'shocks' denote the locations of gas ionized by fast shocks (200-1000 km/s). Grey arrows denote the direction in which gas with a combination of shocks and stellar photoionization, or with a radiative precursor would move (Dopita & Sutherland 1995, Allen et al. 2008, Sharp & Bland-Hawthorne 2010, Rich et al. 2010, 2011). Upper right panel: Co-added nuclear spectrum of BX610, GS3-19791, zC400528, D3a-15504, J0901+1814 and COS4-11363 showing the detection of [OI] λ6300 emission.

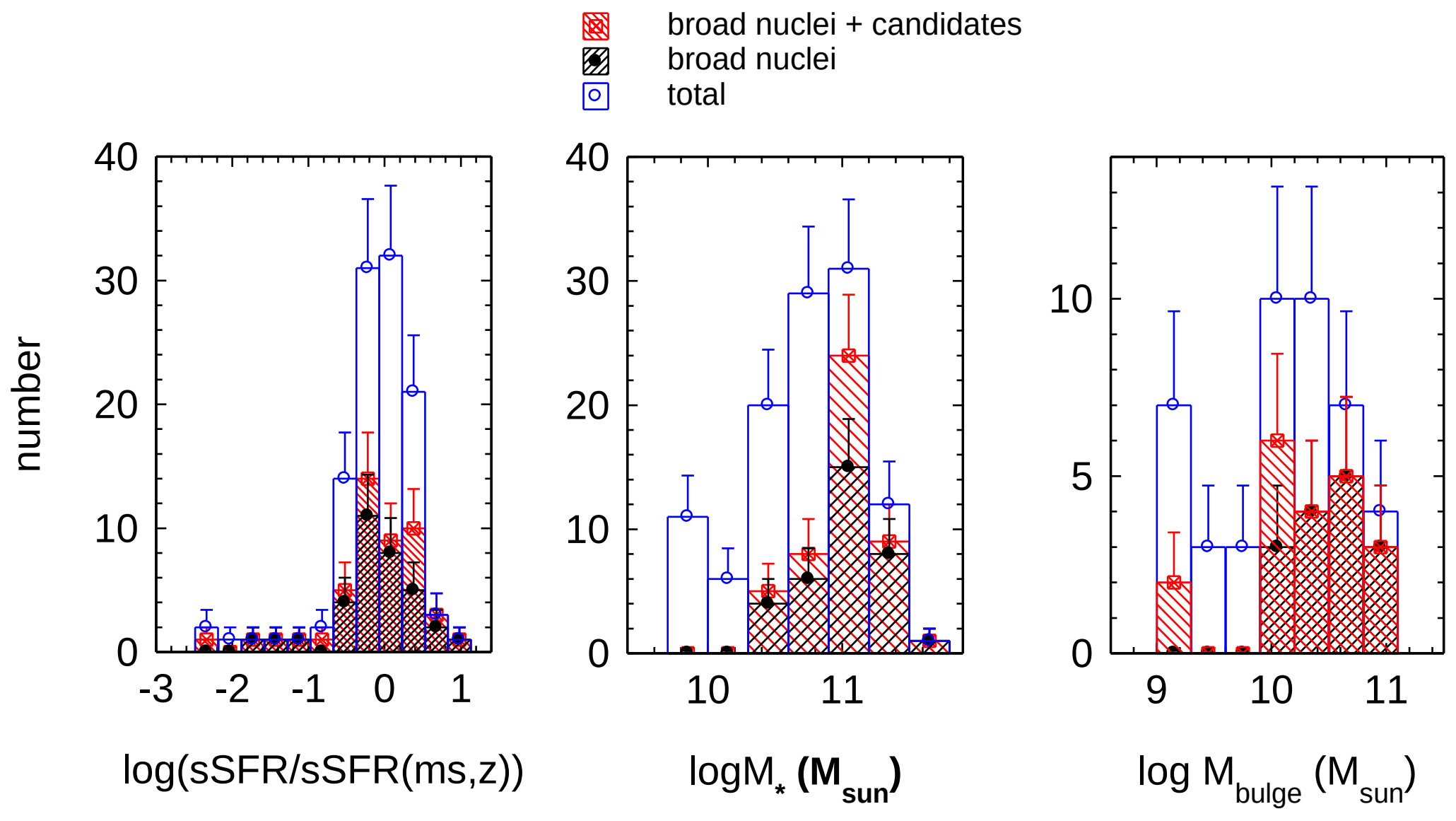

Figure 6. Number of individually detected broad nuclear components for firm detections ("quality 1 +2" in Table 1, black hatched bars, black circles and 1σ errors), for firm plus candidate detections ("quality 0.5+1+2" in Table 1, red hatched bars), compared to all galaxies (blue bars), all as a function of logarithmic offset from the normalized main sequence line (left panel), as a function of total galaxy stellar mass (central panel) and of bulge stellar mass (right panel). The error bars are Poissonian.

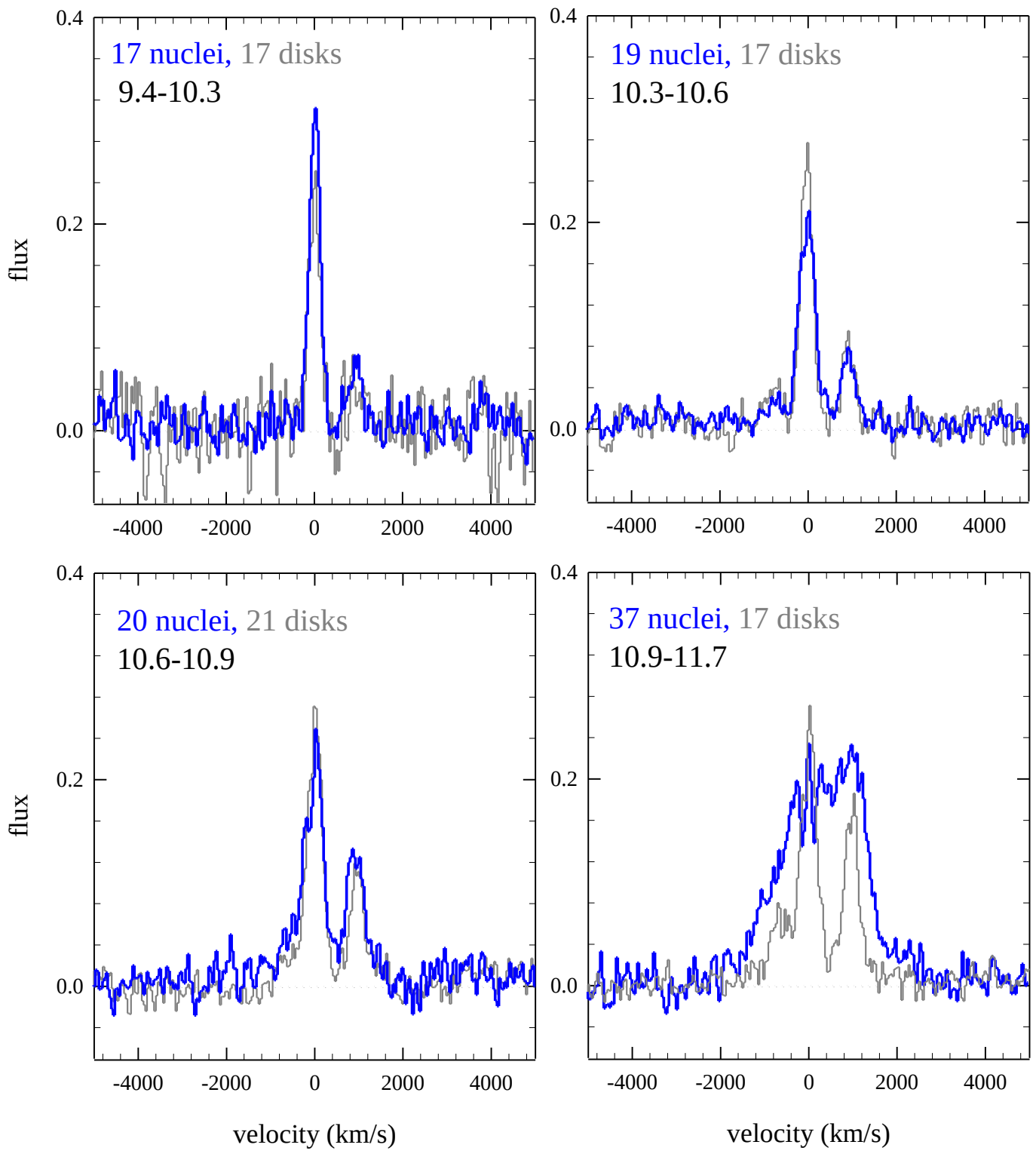

Figure 7: Properties of residual broad component spectra from SNR weighted stacking of all spectra in each of the four mass bins, after removal of the narrow component, as in Figure 4. In each panel the blue spectrum is the broad nuclear residual profile, while the grey spectrum is the outer disk broad residual profile in the same mass bin.

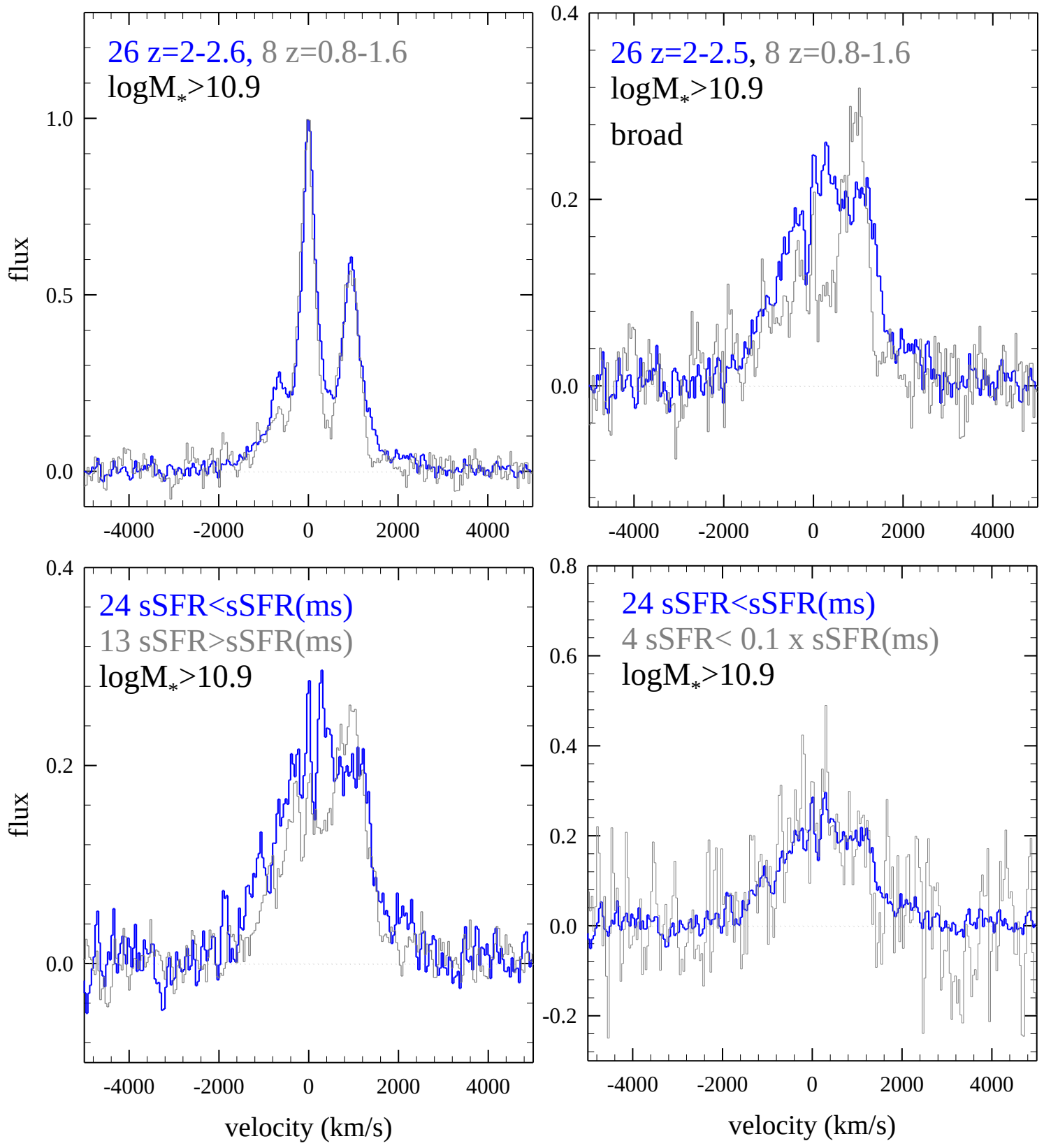

Figure 8: Top: Comparison of broad nuclear spectra (individually detected, including candidates but excluding the BLR sources) at z=2-2.6 (blue) and z=0.8-1.6 (grey), in the mass bin $\log(M_*/M_\odot)\geq 10.9$. The left panel compares the total co-added spectra (weighted by SNR), while the right panel shows the broad components, after removal of the narrow components, as in Figures 4 and 7. Bottom left: Comparison of the weighted, co-added spectra in the $\log(M_*/M_\odot)\geq 10.9$ bin, below (blue) and above (grey) the main sequence. Bottom right: comparison of the broad residual spectrum of the 4 galaxies with a tenth or less the specific star formation rate of the main sequence (grey) with the near- but below-main sequence stack from the bottom left panel (blue).

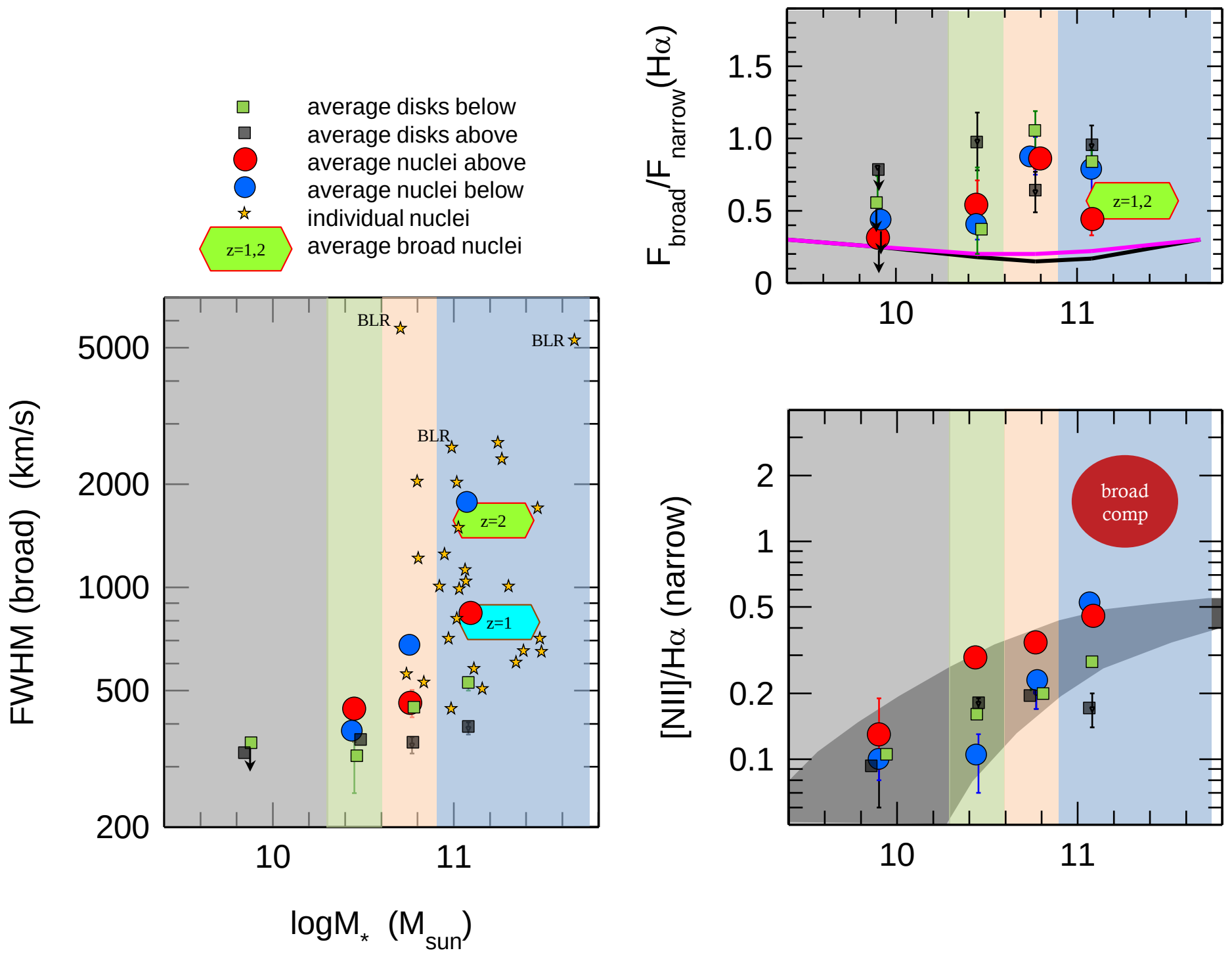

Figure 9. FWHM line width of the broad component (left), narrow and broad [NII] λ6583/Hα flux ratio (bottom right) and broad to narrow Hα flux ratio (top right) of the nuclear and disk spectra, as a function of stellar mass. Filled green and black squares denote weighted stacks in the outer disks, above and below the main sequence line respectively, in the four stellar mass bins marked by grey, green pink and blue shading (same as in Figure 1). Filled blue and red circles show the stacks for the nuclear regions, again above and below the main sequence line. Asterisks denote individual SFGs. Hexagons mark average of the z=0.8-1.6 and 2-2.6 SFGs. The dark grey shading in the lower right panel shows the z~1-2 mass metallicity relation (Erb et al. 2006, Liu et al. 2008, Zahid et al. 2014, E.Wuyts et al. 2014a). The large brown oval marked 'broad comps' and the green hexagon show the ratios of the broad λ 6583 [NII]/Hα lines, while all other symbols refer to the narrow component. The thick black and pink near-horizontal curves in the upper right panel denote the limits of detecting and correctly inferring the width and amplitude relative to the narrow component for a FWHM 500 and 1500 km/s broad emission component in the different stacks (same as right panel in Figure 3).

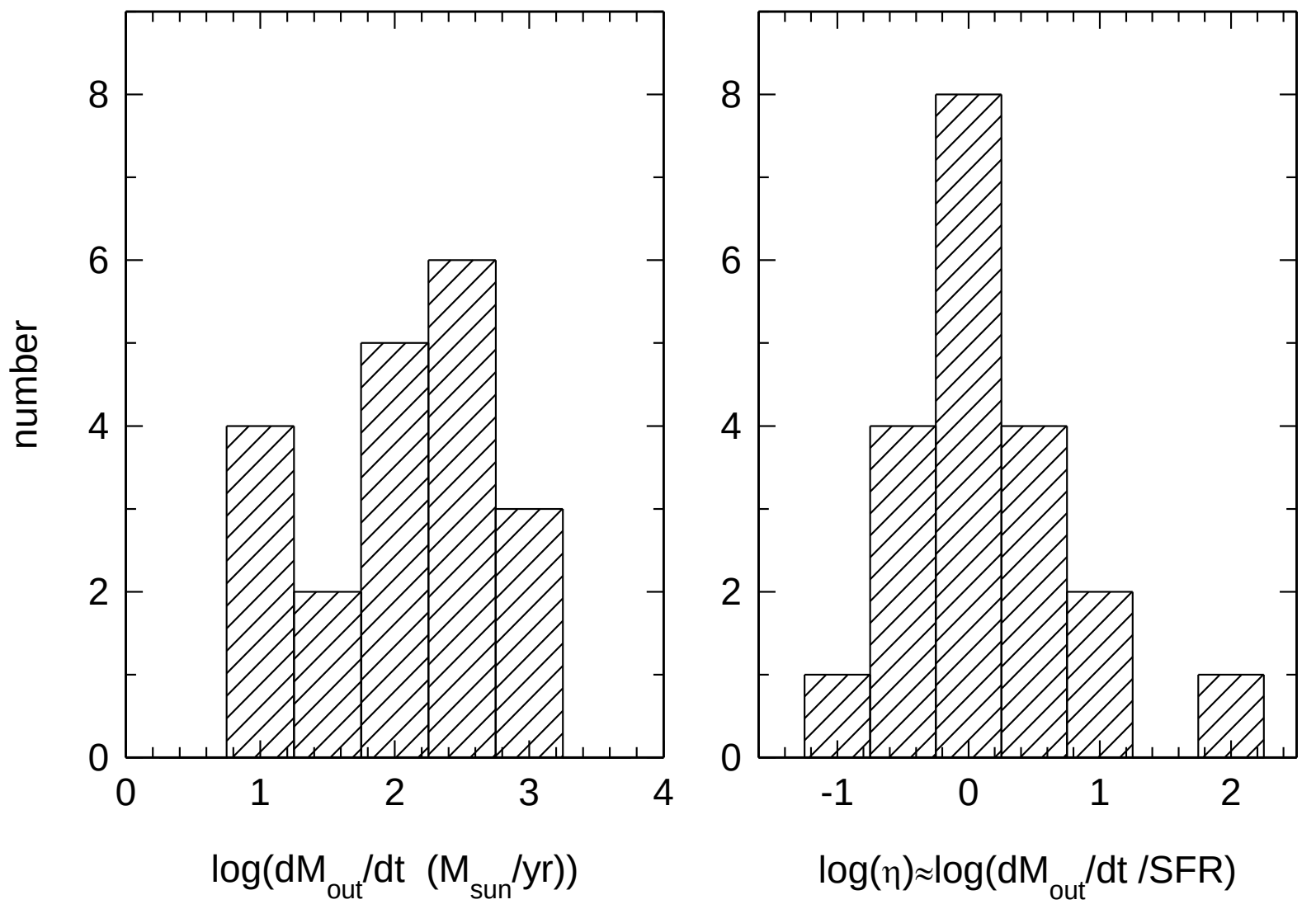

Figure 10. Inferred distribution of mass outflow rates (left) and nuclear mass loading factors (ratio of outflow rate to star formation rate, right) inferred from the data in the 20 $\log M_* > 10.8$ SFGs with good individual broad detections (excluding those SFGs with broad line regions). See section 3.4 and Table 4 for details.

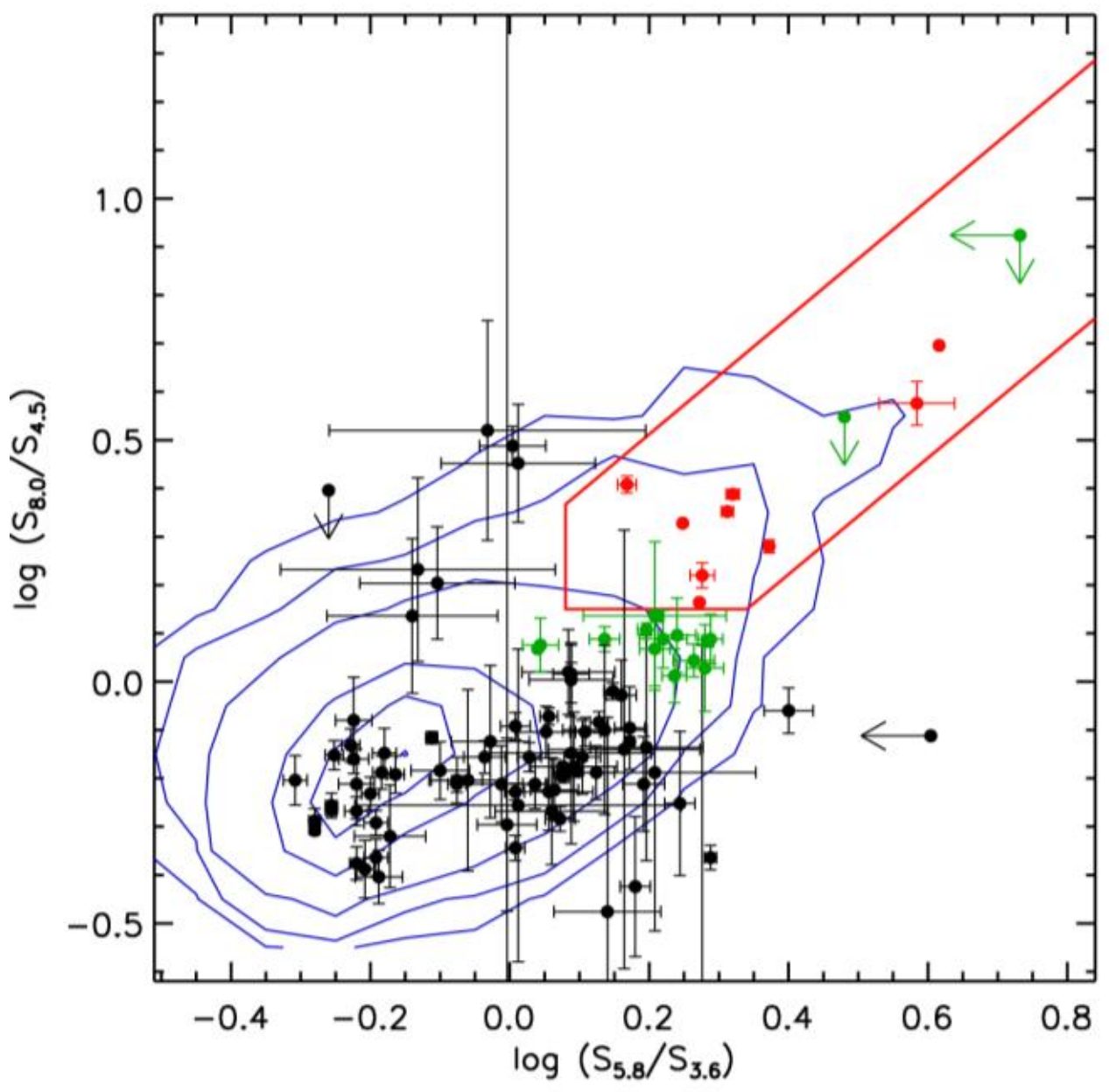

Figure 11. IRAC flux ratio plot adapted from Figure 12 of Donley et al. (2012). The blue contours in the background represent the distribution of 0.7<z<2.6 galaxies based on the COSMOS IRAC catalog (SCOSMOS). Galaxies from our sample with IRAC photometry are plotted as filled black circles with error bars. The Donley et al. (2012) selection box for AGN is shown in red and confirmed AGN in our sample are plotted as red points. Lower-quality candidates are plotted as green points. Only a few galaxies from our sample lie within the Donley et al. (2012) region. Another small number lie close to its boundary and may contain weak AGN.

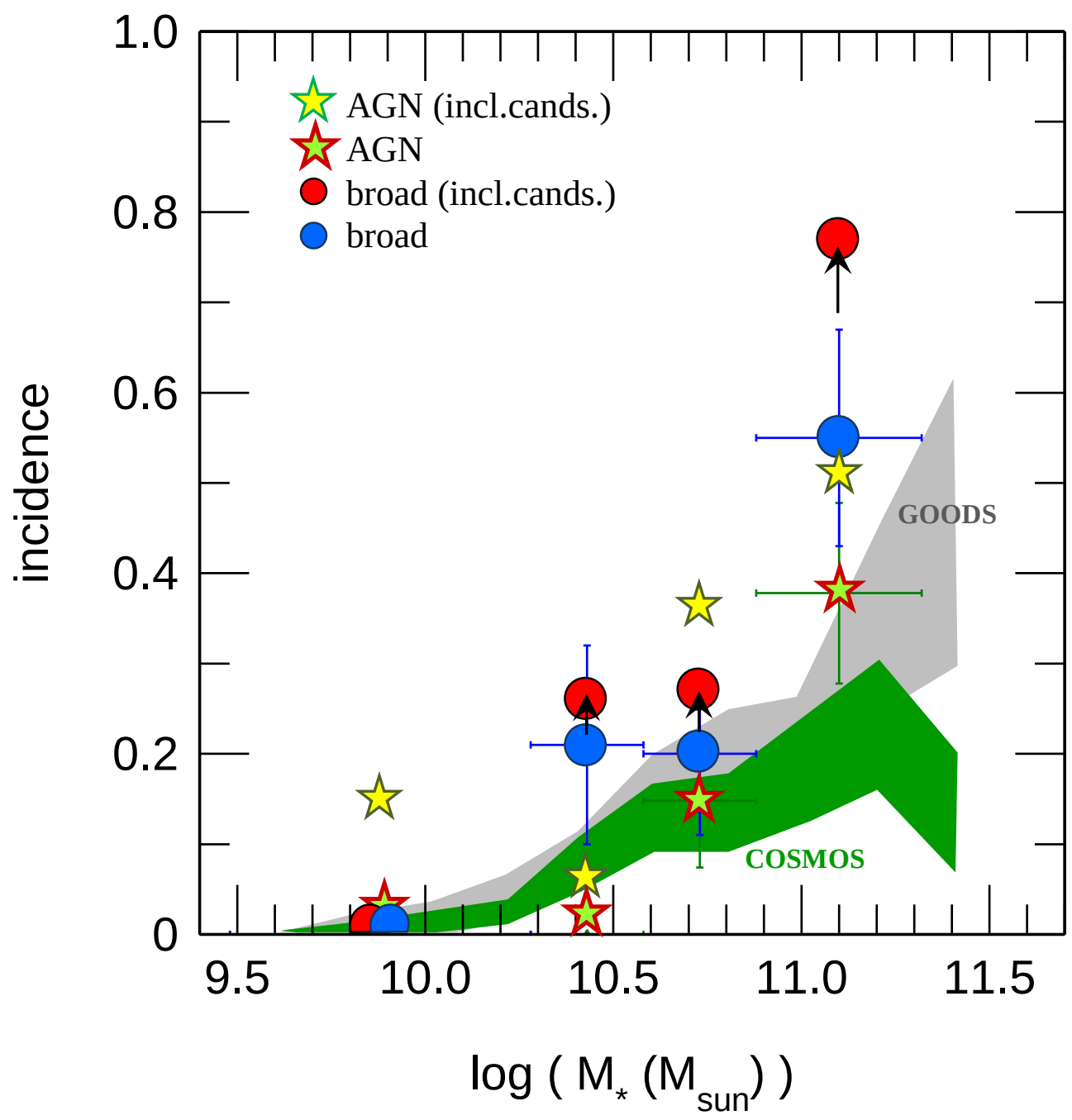

Figure 12. Comparison of the mass dependence of the incidence of broad nuclear emission and AGN identified on the basis of X-ray/optical/infrared and radio criteria. Filled blue circles denote the incidence of firm broad nuclear component detections, and upward pointing arrows ending at the filled red circles show the incidence of the firm plus candidate broad nuclear component detections. Green/brown and yellow/green asterisks denote the incidence of firm AGNs, and AGNs including candidates in our sample (i.e., from the "common sample" described in Section 4.2.1). The grey- and green-shaded distributions denote the AGN incidence as a function of stellar mass as probed in the entire GOODS N/S and COSMOS fields (see Section 4.2).

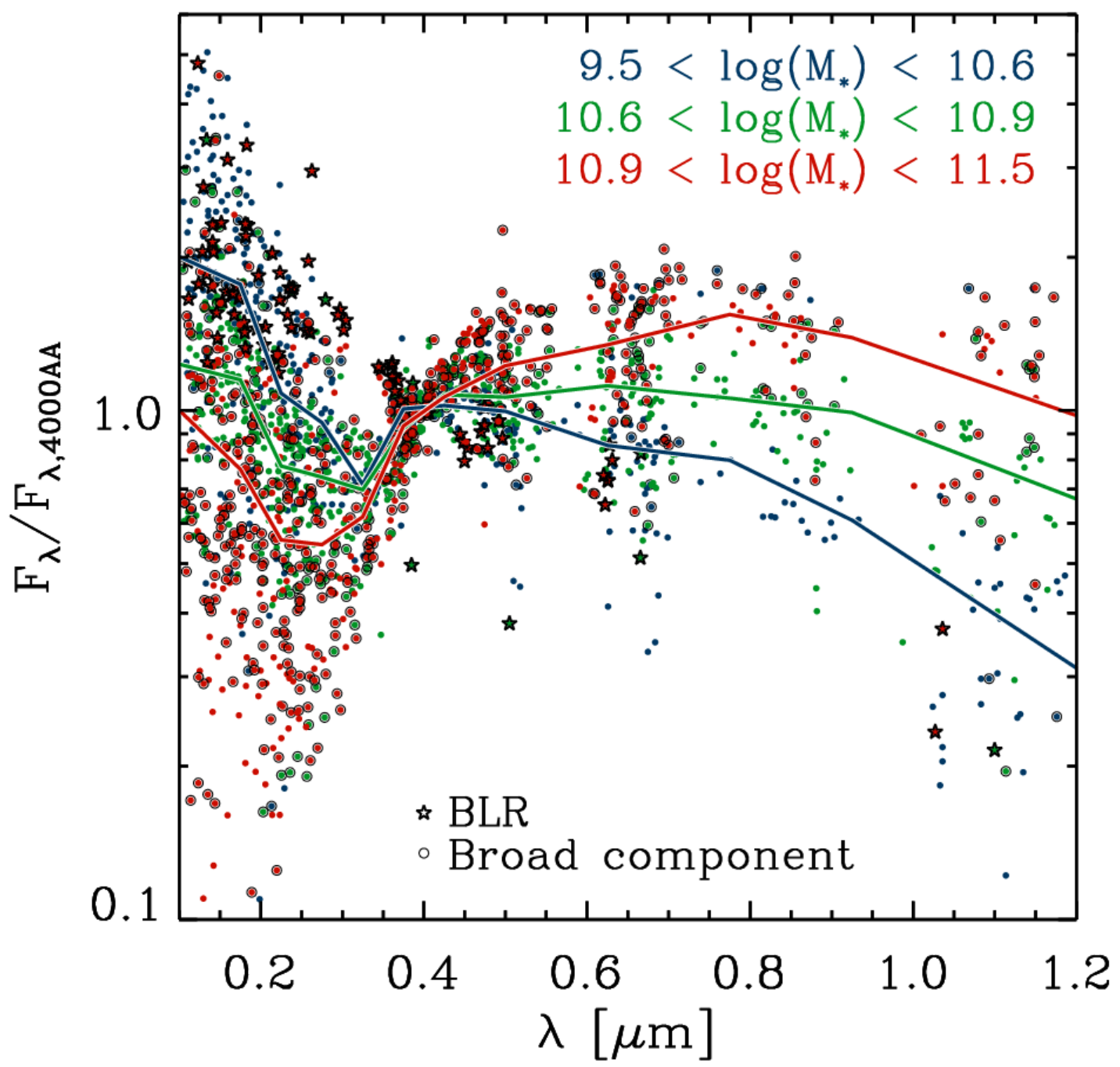

Figure 13. Rest-frame normalized spectral energy distributions of the galaxies in our sample, color-coded by their stellar mass bin as labeled in the plot. Symbols correspond to the photometry of individual galaxies, and thick lines show the median SEDs of galaxies in the three mass bins. Galaxies with a broad component are indicated with a black circle. The photometry of the 3 BLR sources is marked with star symbols. The BLR sources have blue SEDs, presumably due to significant contributions from nuclear emission, complicating estimates of their stellar mass content.

Galaxies featuring broad outflow components on the other hand have stellar SEDs with well-pronounced Balmer 4000Å breaks, confirming their inferred high masses.

## Table 1. SFG sample

| Source | Survey | Kinematics[a] | Mode[b] | $T_{int}$ [c] (h) | SNR center | Broad[d] | $z$ | $\log(M_*/M_\odot)$ | sSFR/sSFR(ms)[e] | [NII]/Hα center | $R_{1/2}$ (kpc) | AGN | $\log(L(AGN)/\mathrm{erg\ s^{-1}})$ [f] | L(SFR)/L(AGN) [g] |
|---|---|---|---|---|---|---|---|---|---|---|---|---|---|---|
| SSA22a-MD41 | SINS/zC-SINF | disk | ss | 7 | 16 | ... | 2.17 | 9.89 | 3.131 | 0.12 | 5.1 | ... | ... | ... |
| ZC-405501 | SINS/zC-SINF | disk | AO | 5.7 | 9 | ... | 2.15 | 9.92 | 1.552 | 0.08 | 7.7 | ... | ... | ... |
| ZC-413507 | SINS/zC-SINF | disk | AO | 5.8 | 8 | ... | 2.48 | 9.94 | 1.335 | 0.10 | 3.6 | cand | ... | ... |
| ZC-405226 | SINS/zC-SINF | disk | AO | 12.3 | 10 | ... | 2.29 | 9.96 | 2.003 | 0.33 | 4.4 | ... | ... | ... |
| ZC-413597 | SINS/zC-SINF | disp | AO | 5.8 | 9 | ... | 2.44 | 9.87 | 1.287 | 0.11 | 2.5 | ... | ... | ... |
| ZC-415876 | SINS/zC-SINF | disk | AO | 5.8 | 11 | ... | 2.44 | 9.96 | 1.050 | 0.12 | 1.9 | ... | ... | ... |
| GMASS-2438 | SINS/zC-SINF | disk | ss | 3.7 | 6 | ... | 1.62 | 10.25 | 2.327 | 0.39 | 8.0 | ... | ... | ... |
| Q2346-BX482 | SINS/zC-SINF | disk | AO | 12.3 | 6 | ... | 2.26 | 10.26 | 1.151 | 0.26 | 5.5 | ... | ... | ... |
| Q1623-BX502 | SINS/zC-SINF | disk | AO | 6.3 | 30 | ... | 2.16 | 9.36 | 0.905 | 0.05 | 1.3 | ... | ... | ... |
| ZC-411737 | SINS/zC-SINF | disk | AO | 4.2 | 8 | ... | 2.44 | 9.54 | 0.781 | 0.06 | 3.1 | cand | ... | ... |
| ZC-410123 | SINS/zC-SINF | disk-disp | AO | 2 | 7 | ... | 2.20 | 9.62 | 0.919 | 0.10 | 4.8 | ... | ... | ... |
| ZC-410041 | SINS/zC-SINF | disk | AO | 6 | 7 | ... | 2.45 | 9.66 | 0.868 | 0.05 | 5.0 | ... | ... | ... |
| ZC-401925 | SINS/zC-SINF | disp | AO | 3.5 | 12 | ... | 2.14 | 9.76 | 0.898 | 0.08 | 2.5 | ... | ... | ... |
| Q1623-BX455 | SINS/zC-SINF | disk | AO | 3.5 | 11 | ... | 2.41 | 10.01 | 0.553 | 0.20 | 2.0 | ... | ... | ... |
| U3-10523 | $KMOS^{3D}$ | disp | s | 7.1 | 25 | ... | 2.16 | 10.05 | 0.210 | 0.18 | 1.2 | ... | ... | ... |
| U3-15027 | $KMOS^{3D}$ | disp | s | 7.1 | 10 | ... | 2.29 | 10.17 | 0.244 | 0.29 | 2.7 | ... | ... | ... |
| GMASS-2540 | SINS/zC-SINF | disk | AO | 10 | 11 | ... | 1.61 | 10.28 | 0.535 | 0.29 | 11.2 | ... | ... | ... |
| ZC-412369 | SINS/zC-SINF | disp | AO | 4 | 28 | 1 | 2.03 | 10.34 | 1.584 | 0.22 | 3.8 | ... | ... | ... |
| SA12-6339 | SINS/zC-SINF | disp | AO | 7.8 | 40 | 2 | 2.30 | 10.41 | 4.307 | 0.18 | 1.6 | ... | ... | ... |
| ZC-407302 | SINS/zC-SINF | disk, merger? | AO | 19 | 30 | 0.5 | 2.18 | 10.39 | 4.003 | 0.24 | 4.6 | ... | ... | ... |
| U3-6856 | $KMOS^{3D}$ | disk | ss | 7 | 11 | ... | 2.30 | 10.41 | 1.027 | 0.21 | 1.9 | ... | ... | ... |
| COS3-21583 | $KMOS^{3D}$ | disk | ss | 1.7 | 20 | ... | 0.89 | 10.50 | 2.887 | 0.25 | 4.4 | ... | ... | ... |
| COS3-1705 | $KMOS^{3D}$ | disk | ss | 3.7 | 50 | ... | 0.83 | 10.55 | 2.502 | 0.35 | 7.5 | ... | ... | ... |
| GS3-24369 | KMOS$^{3D}$ | disk | s | 8.2 | 27 | ... | 0.89 | 10.59 | 2.245 | 0.43 | 1.9 | ... | ... | ... |
| GMASS-2363 | SINS/zC-SINF | disk | AO | 13.7 | 10 | ... | 2.45 | 10.34 | 0.803 | 0.14 | 2.4 | ... | ... | ... |
| COS4-5094 | KMOS$^{3D}$ | disk | ss | 11.3 | 13 | ... | 2.17 | 10.38 | 0.887 | 0.27 | 5.1 | ... | ... | ... |
| U3-10584 | KMOS$^{3D}$ | disk | ss | 7 | 18 | ... | 2.24 | 10.37 | 0.771 | 0.18 | 4.7 | cand | ... | ... |
| GS3-26790 | KMOS$^{3D}$ | disk | ss | 8.9 | 17 | ... | 2.23 | 10.39 | 0.217 | 0.08 | 4.4 | ... | ... | ... |
| U3-3856 | KMOS$^{3D}$ | disk | ss | 4.5 | 10 | ... | 0.80 | 10.40 | 0.565 | 0.38 | 4.7 | ... | ... | ... |
| U3-27143 | KMOS$^{3D}$ | disk | ss | 7 | 25 | 2 | 2.26 | 10.42 | 0.353 | 0.22 | 1.6 | ... | ... | ... |
| GS3-26192 | KMOS$^{3D}$ | disk-disp | s | 8.9 | 25 | ... | 2.32 | 10.45 | 0.475 | 0.10 | 2.6 | ... | ... | ... |
| COS4-15813 | KMOS$^{3D}$ | disk | ss | 8.2 | 20 | ... | 2.36 | 10.57 | 0.612 | 0.10 | 2.5 | ... | ... | ... |
| COS4-4453 | KMOS$^{3D}$ | disp | s | 11.3 | 6 | ... | 2.44 | 10.56 | 0.239 | 0.34 | 3.1 | ... | ... | ... |
| K20-ID8 | SINS/zC-SINF | disk | ss | 3.7 | 23 | ... | 2.22 | 10.51 | 0.622 | 0.29 | 6.0 | ... | ... | ... |
| GS3-22466 | KMOS$^{3D}$ | disk | ss | 8.9 | 8 | ... | 2.23 | 10.56 | 0.923 | 0.28 | 3.9 | ... | ... | ... |
| GS3-27242 | KMOS$^{3D}$ | disk | s | 8.2 | 2 | 1 | 1.03 | 10.58 | 0.856 | 0.51 | 2.6 | ... | ... | ... |
| Q2343-BX389 | SINS/zC-SINF | disk | AO | 5 | 15 | ... | 2.17 | 10.61 | 1.067 | 0.20 | 6.8 | ... | ... | ... |
| ZC-406690 | SINS/zC-SINF | disk | AO | 10 | 6 | 0.5 | 2.20 | 10.62 | 2.508 | 0.27 | 5.5 | ... | ... | ... |
| ZC-403741 | SINS/zC-SINF | disk | AO | 4 | 22 | ... | 1.45 | 10.65 | 1.339 | 0.53 | 2.5 | ... | ... | ... |
| K20-ID7 | SINS/zC-SINF | disk | AO | 7.2 | ... | ... | 2.22 | 10.60 | 1.174 | 0.22 | 8.4 | ... | ... | ... |
| COS3-23443 | KMOS$^{3D}$ | disk | ss | 1.7 | 2 | 0.5 | 0.89 | 10.77 | 2.114 | 0.80 | 5.9 | ... | ... | ... |
| COS3-16954 | KMOS$^{3D}$ | disk | ss | 9.2 | 16 | ... | 1.03 | 10.78 | 3.123 | 0.76 | 7.1 | ... | ... | ... |
| COS3-25038 | KMOS$^{3D}$ | disk | ss | 1.7 | 15 | ... | 0.85 | 10.80 | 1.573 | 0.37 | 25.0 | ... | ... | ... |
| GS3-18419 | KMOS$^{3D}$ | disk | ss | 8.9 | 14 | 2 | 2.31 | 10.81 | 6.975 | 0.70 | 2.8 | det | <45.3 | 18 |
| COS4-4519 | KMOS$^{3D}$ | disk | ss | 11.3 | 20 | ... | 2.23 | 10.61 | 1.729 | 0.30 | 2.4 | cand | ... | ... |
| COS3-18434 | KMOS$^{3D}$ | disk | ss | 3.7 | 20 | ... | 0.91 | 10.82 | 2.014 | 0.46 | 4.3 | ... | ... | ... |
| COS4-19680 | KMOS$^{3D}$ | disk | ss | 8.2 | ... | ... | 2.17 | 10.85 | 1.415 | 0.55 | 2.4 | ... | ... | ... |
| COS4-10347 | KMOS$^{3D}$ | disk | ss | 19.8 | 11 | ... | 2.06 | 10.85 | 1.171 | 0.39 | 4.0 | cand | ... | ... |
| COS3-4796 | KMOS$^{3D}$ | disk | ss | 3.7 | 5 | ... | 1.03 | 10.83 | 1.805 | 0.42 | 6.5 | ... | ... | ... |
| ECDFS-10525 | GNIRS+SINFONI | ... | s | 3 | 2 | 1 | 2.02 | 10.72 | 1.296 | 0.50 | ... | det | <45.6 | 1.3 |
| U3-8493 | KMOS$^{3D}$ | disk | ss | 4.5 | 12 | ... | 0.79 | 10.64 | 0.203 | 0.57 | 2.4 | ... | ... | ... |
| GS3-24364 | KMOS$^{3D}$ | disk-disp | ss | 8.9 | 25 | ... | 2.33 | 10.70 | 0.497 | 0.17 | 5.3 | ... | ... | ... |
| COS4-13701 | KMOS$^{3D}$ | disk | ss | 8.2 | 25 | ... | 2.17 | 10.67 | 0.991 | 0.23 | 4.0 | ... | ... | ... |
| COS3-11468 | KMOS$^{3D}$ | ... | s | 4.2 | weak | ... | 0.89 | 10.83 | 0.288 | no | 3.7 | ... | ... | ... |
| Q1623-BX599 | SINS/zC-SINF | disk, merger? | AO | 2 | 25 | 1 | 2.33 | 10.75 | 0.511 | 0.17 | 3.1 | ... | ... | ... |
| Q1623-BX663 | SINS/zC-SINF | disk | AO-s | 8.8 | 15 | 2 | 2.43 | 10.81 | 0.664 | 0.43 | 6.5 | det | <46.0 | 0.35 |
| U3-25105 | KMOS$^{3D}$ | disk | ss | 7 | 12 | 1 | 2.29 | 10.85 | 0.826 | 0.50 | 6.0 | cand | ... | ... |
| U3-13321 | KMOS$^{3D}$ | disk | ss | 4 | 2 | ... | 0.91 | 10.85 | 0.515 | 0.82 | 3.6 | ... | ... | ... |
| GOODSN-19394 | LUCI | ... | s | 4 | ... | ... | 1.45 | 10.7 | 0.193 | 0.19 | 21.0 | ... | ... | ... |
| GOODSN-31720 | LUCI | ... | s | 4 | ... | ... | 2.48 | 10.7 | 0.272 | 0.23 | ... | cand | ... | ... |
| GOODSN-03493 | LUCI | ... | s | 4 | ... | ... | 2.46 | 10.8 | 0.364 | 0.40 | ... | cand | ... | ... |
| GOODSN-07923 | LUCI | BLR | s | 4 | ... | 2 | 2.24 | 10.7 | 0.425 | broad | ... | det | 45.6 | 0.5 |
| 1030-807 | GNIRS+SINFONI | ... | s | 3 | ... | ... | 2.37 | 10.81 | 0.004 | 0.33 | 3.3 | ... | ... | ... |
| ECDFS-5754 | GNIRS+SINFONI | ... | s | 3 | ... | ... | 2.04 | 10.81 | 0.989 | 0.20 | 5.5 | ... | ... | ... |
| COS4-18859 | KMOS$^{3D}$ | ... | s | 8.2 | ... | ... | 2.61 | 10.75 | 0.720 | no | 0.8 | ... | ... | ... |
| COS4-16342 | KMOS$^{3D}$ | disk | s | 8.2 | 8 | ... | 2.47 | 10.85 | 0.772 | 0.27 | 5.1 | cand | ... | ... |
| COS4-4717 | KMOS$^{3D}$ | disk | s | 11.3 | 9 | ... | 2.44 | 10.93 | 1.916 | 0.36 | 4.1 | cand | ... | ... |
| ZC-400528 | SINS/zC-SINF | disk+merger | AO | 4 | 20 | 2 | 2.39 | 11.04 | 1.768 | 0.75 | 2.0 | ... | ... | ... |
| D3a-6397 | SINS/zC-SINF | disk | AO | 8.5 | 18 | 2 | 1.50 | 11.08 | 5.052 | 0.77 | 6.0 | ... | ... | ... |
| ZC-400569 central disk | SINS/zC-SINF | disk+merger | AO | 22 | 18 | 1 | 2.24 | 11.08 | 1.213 | 0.73 | 6.4 | ... | ... | ... |
| GS3-31118 | KMOS$^{3D}$ | disk | ss | 4.4 | 8 | 0.5 | 2.45 | 11.13 | 1.898 | 1.49 | 1.1 | cand | ... | ... |
| U3-16262 | KMOS$^{3D}$ | disk | ss | 5.8 | 11 | 0.5 | 2.30 | 11.18 | 1.642 | 0.63 | 2.5 | cand | ... | ... |
| GS3-19791 (K20-ID5) | KMOS$^{3D}$ | disk | ss | 4.4 | 30 | 2 | 2.22 | 11.31 | 1.649 | 0.80 | 3.6 | det | 44.6 | 29 |
| D3a-6004 | SINS/zC-SINF | disk | AO | 4.7 | 9 | 2 | 2.39 | 11.50 | 1.446 | 1.09 | 5.0 | ... | ... | ... |
| J0901+1814 | SINFONI | disk | AO | 9 | 10 | 2 | 2.26 | 11.49 | 2.489 | 0.87 | 2.0 | det | ... | ... |
| EGS13011166 | LUCI | disk | ss | 12 | 8 | 0.5 | 1.53 | 11.04 | 2.367 | 0.56 | 6.0 | det | ... | ... |
| D3a-7144 | SINS/zC-SINF | disk | s | 2 | 9 | 0.5 | 1.65 | 11.07 | 1.565 | 0.87 | 4.6 | det | ... | ... |
| COS4-14596 | KMOS$^{3D}$ | BLR | ss | 8.2 | ... | 2 | 2.44 | 11.68 | 2.503 | broad | 0.2 | det | 45.8 | 5 |
| COS4-13174 | KMOS$^{3D}$ | disk | ss | 19.7 | 15 | 1 | 2.10 | 11.03 | 1.469 | 0.48 | 6.4 | cand | ... | ... |
| COS4-10056 | KMOS$^{3D}$ | disk | s | 19.7 | 5 | ... | 2.56 | 11.03 | 1.096 | 0.50 | 4.5 | ... | ... | ... |
| COS4-21492 | KMOS$^{3D}$ | BLR | s | 8.1 | ... | 2 | 2.47 | 11.00 | 2.707 | broad | 0.4 | det | 46.1 | 1.4 |
| COS4-6963 | KMOS$^{3D}$ | merger? | s | 11.3 | 8 | 2 | 2.30 | 10.96 | 0.059 | 0.20 | 2.2 | ... | ... | ... |
| GS3-21045 | KMOS$^{3D}$ | disk | ss | 8.2 | 12 | ... | 0.96 | 10.92 | 0.506 | 0.85 | 9.2 | ... | ... | ... |
| GS3-22005 | KMOS$^{3D}$ | disk | ss | 8.2 | 10 | 1 | 0.95 | 10.93 | 0.410 | 0.53 | 32.0 | ... | ... | ... |
| U3-12280 | KMOS$^{3D}$ | disk+merger | ss | 8.9 | 7 | 2 | 1.03 | 10.98 | 0.516 | 0.83 | 4.1 | ... | ... | ... |
| Q2343-BX610 | SINS/zC-SINF | disk | AO | 8.3 | 22 | 2 | 2.21 | 11.00 | 0.548 | 0.58 | 8.0 | ... | ... | ... |
| U3-15226 | KMOS$^{3D}$ | disk | ss | 8.9 | 10 | ... | 0.92 | 11.00 | 0.856 | 0.85 | 5.8 | ... | ... | ... |
| D3a-15504 | SINS/zC-SINF | disk | AO | 23 | 30 | 2 | 2.38 | 11.04 | 0.949 | 0.48 | 6.7 | det | ... | ... |
| GS3-28464 | KMOS$^{3D}$ | disk | ss | 17 | 3 | 0.5 | 2.30 | 11.04 | 0.409 | 0.54 | 1.9 | det | 44.4 | 10 |
| GS3-25445 | KMOS$^{3D}$ | disk | ss | 4.4 | 12 | 0.5 | 2.43 | 11.13 | 0.744 | 0.55 | 0.7 | ... | ... | ... |
| COS3-644 | KMOS$^{3D}$ | disk | ss | 3.7 | 10 | 1 | 0.88 | 11.17 | 0.484 | 0.99 | 5.0 | ... | ... | ... |
| COS3-8390 | KMOS$^{3D}$ | disk | ss | 3.7 | 2 | ... | 0.98 | 11.27 | 0.505 | 1.00 | 3.8 | ... | ... | ... |
| U3-23710 | KMOS$^{3D}$ | disk | ss | 7.1 | 10 | 2 | 2.53 | 11.03 | 0.309 | 0.59 | 4.7 | ... | ... | ... |
| GS3-28008 | KMOS$^{3D}$ | nucleus only | ss | 17 | 4 | 2 | 2.29 | 11.36 | 0.493 | 0.87 | 3.3 | det | 45.9 | 0.5 |
| GS3-7562 | KMOS$^{3D}$ | disk | ss | 7.5 | 2 | 0.5 | 2.04 | 11.32 | 0.670 | 0.20 | 6.5 | ... | ... | ... |
| GOODSN-29999 | LUCI | ... | s | 4 | ... | ... | 1.53 | 11 | 0.493 | 0.40 | ... | ... | ... | ... |
| GOODSN-22747 | LUCI | ... | s | 4 | 5 | 2 | 1.45 | 11 | 0.214 | 1.30 | ... | det | 45.9 | 0.08 |
| GOODSN-22412 | LUCI | ... | s | 4 | 5.3 | 0.5 | 1.52 | 11 | 0.259 | 0.30 | ... | ... | ... | ... |
| Q2343-BX442 | LUCI | disk | s | 4 | ... | ... | 2.18 | 11.1 | 0.256 | OH | ... | ... | ... | ... |
| GOODSN-17020 | LUCI | ... | s | 4 | ... | 0.5 | 2.33 | 11.1 | 0.154 | 1.20 | ... | det | 44.6 | 2.6 |
| 1030-1531 | GNIRS+SINFONI | ... | s | 3 | ... | ... | 2.61 | 11 | 0.765 | 0.35 | 3.9 | ... | ... | ... |
| 1030-2026 | GNIRS+SINFONI | ... | s | 3.1 | 4 | 2 | 2.51 | 11.25 | 0.033 | 0.64 | 1.5 | det | ... | ... |
| 1030-2329 | GNIRS+SINFONI | ... | s | 3 | 3 | 1 | 2.24 | 10.95 | 0.020 | 0.71 | 1.3 | ... | ... | ... |
| 1030-2728 | GNIRS+SINFONI | ... | s | 2 | 1 | 0.5 | 2.50 | 11.18 | 0.003 | 0.63 | 1.0 | ... | ... | ... |
| ECDFS-3662 | GNIRS+SINFONI | ... | s | 3 | 2 | 1 | 2.35 | 11.09 | 0.316 | 0.56 | 1.7 | ... | ... | ... |
| ECDFS-3694 | GNIRS+SINFONI | ... | s | 4 | ... | ... | 2.12 | 11.36 | 0.541 | 0.45 | 8.6 | ... | ... | ... |
| ECDFS-3896 | GNIRS+SINFONI | ... | s | 2 | 1.5 | 2 | 2.31 | 11.23 | 0.473 | 1.09 | 1.7 | ... | ... | ... |
| COS4-3206 | KMOS$^{3D}$ | disk | ss | 11.5 | 10 | 1 | 2.10 | 11.40 | 0.525 | 0.60 | 6.2 | det | <45.7 | 0.84 |
| COS4-11363 | KMOS$^{3D}$ | merger? | s | 19.7 | 33 | 2 | 2.10 | 11.28 | 0.447 | 0.60 | 2.2 | det | 46.3 | 0.2 |
| COS4-12995 | KMOS$^{3D}$ | disk | s | 19.7 | 1.2 | ... | 2.44 | 11.22 | 0.008 | 1.50 | 1.4 | cand | ... | ... |

[a] Kinematic classification of galaxy from H$\alpha$ data; "disk" stands for rotation, "disp" for dispersion dominated kinematics, "merger" for perturbed motions in a major merger system, and "BLR" for a compact AGN broad line region component.

[b] Observing mode for the data used in this work. “AO” indicates adaptive optics-assisted observations with FWHM resolution of 0.2″–0.3″; “s” and “ss” indicate seeing-limited observations with FWHM resolution of 0.5″–0.7″ (“ss” denotes objects for which the kinematics are well resolved).

[c] Total on-source integration time of the observations.

[d] Identification of a broad nuclear emission component: “2” for a strong nuclear broad component, “1” for a clear nuclear broad component, “0.5” for a candidate nuclear broad component.

[e] Specific SFR normalized to that of the main sequence of SFGs at the redshift and stellar mass of each object using the parametrization of Whitaker et al. (2012), applicable for $\log(M_*/M_\odot) > 10$.

[f] The bolometric AGN luminosity is estimated either from the absorption corrected X-ray luminosity (as in Rosario et al. 2012), or from the rest-frame 8μm-luminosity of power-law mid-IR SEDs extrapolated to the total blue bump luminosity with AGN SEDs (Richards et al. 2006), or an average. If only a mid-IR estimate is available, we consider this luminosity an upper limit to the AGN luminosity.

[g] Ratio of the AGN to galaxy integrated star formation rate luminosity (assuming $L(SFR)=1\times10^{10}$ x SFR).

**Table 2. Spectral Properties of log(M/M$_{\odot}$)>10.9 stacks**

| Property | narrow component | broad component |
|---|---|---|
| $\Delta v$ (FWHM) (km/s) | center (31 objects): 365 (6)<br><br>disk (16 objects): 160 (4) | 1711 (70)<br><br>440 (30) |
| $\delta v_{broad}$(km/s)[a] | center: -<br><br>disk: - | -130 (40)<br><br>11 (8) |
| F(H$\alpha$)$_{broad}$ / F(H$\alpha$)$_{narrow}$[b] | center: -<br><br>disk: - | 0.4 (0.1)<br><br>0.85 (0.1) |
| F([NII] $\lambda$6583)/F(H$\alpha$) | center: 0.55 (0.13)<br><br>disk: 0.23 (0.02) | 2.7 (0.7)<br><br>0.7 (0.06) |
| F([SII] $\lambda\lambda$6716+6731) / F(H$\alpha$)$_{narrow}$ | center: 0.27 (0.03)<br><br>disk: 0.2 (0.03) | 0.2 (0.03)<br><br>0.12 (0.03) |
| F([SII] $\lambda$6716)/F( [SII] $\lambda$6731) | center: 1.07 (0.08)<br><br>disk: 1.13 (0.1) | ~1 |
| F([OI] $\lambda$6300)/F(H$\alpha$)$_{narrow}$ | 0.099 (0.025) for best 6 | - |
| F(5007 [OIII]/F(H$\beta$)$_{narrow}$ | 4 (-1,+4) for best 6 | - |

NOTE --- Values given in parenthesis are the uncertainties of the measurements.
[a] Velocity offset between the centroid velocity of the broad component relative to the narrow component.
[b] Ratio of the integrated Hα flux in the broad and narrow components.

## Table 3. Incidence of broad nuclear emission components and AGN

| $\log(M_*/M_\odot)$ | number of SFGs | number broad nuclei[a] | number AGN | broad nuclei fraction broad[b] | AGN fraction AGN[c] |
|---|---|---|---|---|---|
| 10.9 – 11.7 | 44 | 24 (34) | 13 (21) | $0.55\ (0.77)_{\pm 0.12}$ | $0.38\ (0.51)_{\pm 0.11}$ |
| 10.6 – 10.9 | 30 | 6 (8) | 5 (9) | $0.2\ (0.27)_{\pm 0.09}$ | $0.15\ (0.37)_{\pm 0.09}$ |
| 10.3 – 10.6 | 19 | 4 (5) | 0 (1) | $0.21\ (0.26)_{0.11}$ | $0\ (0.06)_{\pm 0.06}$ |
| 9.4 – 10.3 | 17 | 0 | 0 (2) | 0 | $0\ (0.15)_{\pm 0.11}$ |

[a] The first number denotes the number of SFGs with broad line components of quality 1 and 2 in Table 1, the number in parentheses denotes the number with quality 1+ 2 + candidates (0.5).
[b] The first number denotes the fraction of SFGs (of the total in that mass bin) of broad line components of quality 1 and 2 in Table 1, the number in parentheses denotes the fraction of quality 1+ 2 + candidates (0.5). The quoted uncertainty in the subscript is the 1σ Poissonian uncertainty.
[c] The first number denotes the fraction of SFGs in the common sample (of the total in that mass bin) that are firmly identified as AGN from at least one of the AGN identifying criteria (X-ray, mid-IR, radio or optical spectroscopy), the number in parentheses denotes the fraction of SFGs in the common sample that either are firm or candidate AGNs. The quoted uncertainty in the subscript is the 1σ Poissonian uncertainty.

## Table 4. Outflow Parameters

| Source | z | $\log(M_*/M_\odot)$ | sSFR/sSFR(ms) | SFR (nucleus) | $R_{HWHM}$ | $F_{broad}/F_{narrow}$ Hα | $L(Ha)_{broad,0}$ | $v_{out}$ | $n(e)_{broad}$ | $M_{broad}$(HII+He) | $dM/dt_{out}$ | $\eta=dM_{out}/dt/SFR$ ionized gas | momentum ratio outflow/radiation | energy ratio dE/dt/L |
|---|---|---|---|---|---|---|---|---|---|---|---|---|---|---|
| - | - | - | - | $(M_\odot/yr)$ | (kpc) | - | (erg/s) | (km/s) | $(cm^{-3})$ | $(M_\odot)$ | $(M_\odot/yr)$ | - | - | - |
| | **1** | **2** | **3** | **4** | **5** | **6** | **7** | **8** | **9** | **10** | **11** | **12** | **13** | **14** |
| Q1623-BX663 | 2.43 | 10.81 | 0.664 | 30 | 1.3 | 1.1 | 6.9E+42 | 1300 | 80 | 2.8E+08 | 288 | 3.1 | 62 | 1.3E-01 |
| U3-25105 | 2.29 | 10.85 | 0.826 | 32 | 1.3 | 0.7 | 4.6E+42 | 214 | 80 | 1.9E+08 | 32 | 0.3 | 1 | 3.8E-04 |
| ZC-400528 | 2.39 | 11.04 | 1.768 | 120 | 1.3 | 0.79 | 2E+43 | 802 | 80 | 8.1E+08 | 513 | 1.7 | 17 | 2.3E-02 |
| D3a-6397 | 1.50 | 11.08 | 5.052 | 73 | 1.3 | 0.6 | 9.2E+42 | 520 | 80 | 3.7E+08 | 153 | 0.3 | 5 | 4.7E-03 |
| ZC-400569 central disk | 2.24 | 11.08 | 1.213 | 92.0 | 1.3 | 0.3 | 5.80E+42 | 350 | 80 | 2.35E+08 | 65 | 0.3 | 1 | 7.2E-04 |
| GS3_19791 (K20-ID5) | 2.22 | 11.31 | 1.649 | 148 | 1.3 | 3.3 | 1.0E+44 | 530 | 80 | 4.2+09 | 1743 | 5.3 | 31 | 2.7E-02 |
| D3a-6004 | 2.39 | 11.50 | 1.446 | 44 | 1.3 | 2.9 | 2.7E+43 | 420 | 80 | 1.1E+09 | 365 | 1 | 17 | 1.2E-02 |
| J0901+1814 | 2.26 | 11.49 | 2.489 | 200 | 1.3 | 1.1 | 4.63E+43 | 323 | 80 | 1.9E+09 | 481 | 0.8 | 4 | 2.1E-03 |
| COS4-13174 | 2.10 | 11.03 | 1.469 | 90 | 1.3 | 0.6 | 1.1E+43 | 350 | 80 | 4.6E+08 | 127 | 0.6 | 2 | 1.4E-03 |
| COS4-6963 | 2.30 | 10.96 | 0.059 | 6.3 | 1.3 | 3 | 4E+42 | 480 | 80 | 1.6E+08 | 61 | 6.8 | 23 | 1.8E-02 |
| GS3-22005 | 0.95 | 10.93 | 0.410 | 2.1 | 1.3 | 0.8 | 3.6E+41 | 700 | 80 | 1.5E+07 | 8 | 0.6 | 13 | 1.5E-02 |
| U3-12280 | 1.03 | 10.98 | 0.516 | 5.7 | 1.3 | 1 | 1.2E+42 | 350 | 80 | 4.8E+07 | 13 | 0.6 | 4 | 2.4E-03 |
| Q2343-BX610 | 2.21 | 11.00 | 0.548 | 13 | 1.3 | 0.2 | 5.6E+41 | 500 | 80 | 2.3E+07 | 9 | 0.1 | 2 | 1.4E-03 |
| D3a-15504 | 2.38 | 11.04 | 0.949 | 24 | 1.3 | 0.7 | 3.5E+42 | 475 | 80 | 1.4E+08 | 54 | 0.3 | 5 | 4.2E-03 |
| COS3-644 | 0.88 | 11.17 | 0.484 | 6.6 | 1.3 | 1 | 1.4E+42 | 300 | 80 | 5.6E+07 | 13 | 0.6 | 3 | 1.5E-03 |
| U3-23710 | 2.53 | 11.03 | 0.309 | 18 | 1.3 | 1.5 | 5.8E+42 | 1300 | 80 | 2.4E+08 | 243 | 4.5 | 85 | 1.8E-01 |
| GS3-28008 | 2.29 | 11.36 | 0.493 | 108 | 1.3 | 1 | 2.3E+43 | 300 | 80 | 9.2E+08 | 219 | 2.0 | 3 | 1.5E-03 |
| 1030-2026 | 2.51 | 11.25 | 0.033 | 7.1 | 1.3 | 10 | 1.5E+43 | 1300 | 80 | 6.0E+08 | 620 | 87.8 | 564 | 1.2E+00 |
| COS43206 | 2.10 | 11.40 | 0.525 | 44 | 1.3 | 1.3 | 1.2E+43 | 300 | 80 | 4.9E+08 | 116 | 1.1 | 4 | 2.0E-03 |
| COS4-11363 | 2.10 | 11.28 | 0.447 | 42 | 1.3 | 2 | 1.7E+43 | 1240 | 80 | 7.1E+08 | 695 | 8.4 | 103 | 2.1E-01 |